\documentclass[a4paper,11pt]{article}
\usepackage{jheppub}
\usepackage[T1]{fontenc}
\usepackage{enumerate}
\usepackage{hyperref}
\usepackage{multirow}
\usepackage{lscape}
\usepackage{tabularx}
\usepackage[tableposition=top]{caption}

\renewcommand{\d}{\mathrm{d}}
\newcommand{\as}{\alpha_s}
\newcommand{\asb}{\bar{\alpha}_s}
\def\X{{\scriptscriptstyle X}}
\def\R{{\scriptscriptstyle\mathrm{R}}}
\def\V{{\scriptscriptstyle\mathrm{V}}}
\def\MC{{\scriptscriptstyle\mathrm{MC}}}
\def\inn{\text{in}}
\def\out{\text{out}}

\def\CF{\mathrm{C_F}}
\def\CFsq{\mathrm{C_F^2}}
\def\CFcub{\mathrm{C_F^3}}
\def\CFfour{\mathrm{C_F^4}}

\def\CA{\mathrm{C_A}}
\def\CAsq{\mathrm{C_A^2}}
\def\CAcub{\mathrm{C_A^3}}
\def\CAfour{\mathrm{C_A^4}}

\def\cA{\mathcal{A}} \def\cAb{\mathcal{\bar{A}}}
\def\cB{\mathcal{B}} 
\def\cC{\mathcal{C}}

\def\cF{\mathcal{F}}
\def\cG{\mathcal{G}}

\def\cN{\mathcal{N}} 
\def\cO{\mathcal{O}}
\def\cQ{\mathcal{Q}}
\def\cS{\mathcal{S}}

\def\cW{\mathcal{W}}
\def\oW{\overline{\mathcal{W}}}

\def\s{\sigma}
\def\S{\Sigma}
\def\O{\Omega}
\def\Ob{\bar{\Omega}}

\def\qb{\bar{q}}
\def\Uh{\hat{\mathcal{U}}}
\def\uh{\hat{u}}

\title{\boldmath Jet-mass in V/H+jet up to four-loops with $k_t$ clustering}

\author[a]{Kamel Khelifa-Kerfa}
\author[b,1]{and Mohamed Benghanem \note{Corresponding author.}}

\affiliation[a]{
Department of Physics, Faculty of Science and Technology\\
University Ahmed Zabana of Relizane, Relizane 48000, Algeria}
\affiliation[b]{
Department of Physics, Faculty of Science\\
Islamic University of Madinah, Madinah 42351, Saudi Arabia}

\emailAdd{kamel.khelifakerfa@univ-relizane.dz}
\emailAdd{mbenghanem@iu.edu.sa}

\abstract{
We extend the work of \cite{Khelifa-Kerfa:2024udm} to the case in which final‑state jets, produced in association with a Higgs or vector boson, are defined using the $k_t$ algorithm. We thereby compute the full distribution of the invariant mass squared of the leading, highest‑$p_T$ jet, including both clustering and non‑global logarithms, up to four-loops in perturbation theory. Our results are derived within the eikonal approximation under the assumption of strong ordering in the momenta of the final‑state partons, and are consequently valid up to single‑logarithmic accuracy. The final semi‑analytical expressions retain the complete dependence on both colour and the jet radius.
The broad features of $k_t$ clustering observed in $e^+e^-$ processes persist in hadronic collisions, together with novel characteristics that are absent in the $e^+e^-$ environment.
}

\keywords{QCD, LHC, Higgs, Jets}

\arxivnumber{}

\begin{document}
\maketitle
\flushbottom

\section{Introduction}
\label{sec:intro}

The study of jet substructure observables, particularly the invariant mass distribution of jets produced in association with heavy bosons, represents a cornerstone of precision QCD at hadron colliders. These observables are highly sensitive to soft and collinear radiation, making them powerful probes of QCD dynamics and essential tools for Standard Model tests and new physics searches. The theoretical description of such observables requires resummation of large logarithmic terms that arise in the small jet-mass limit $\rho = m_j^2/p_{t,j}^2 \ll 1$, where fixed-order perturbative calculations become inadequate.
These logarithmic enhancements originate from multiple sources, including soft and collinear emissions from hard partons (global logarithms), as well as more intricate patterns arising from correlations between emissions in disparate angular regions (non-global logarithms, or NGLs).

The treatment of NGLs presents significant theoretical challenges owing to their non‑abelian nature and sensitivity to the jet definition. Early studies in $e^+e^-$ annihilation demonstrated that NGLs first appear at two-loop order and exhibit a characteristic Sudakov‑suppressed form \cite{Dasgupta:2001sh, Banfi:2002hw, Banfi:2010pa, Khelifa-Kerfa:2011quw, Khelifa-Kerfa:2015mma, Benslama:2020wib}. In hadronic collisions, the complexity increases substantially due to initial‑state radiation and more intricate colour flows \cite{Dasgupta:2012hg}. The choice of jet algorithm plays a crucial role in determining the structure of these logarithms. While the anti‑$k_t$ algorithm \cite{Cacciari:2008gp} is widely employed for its simplicity and infrared safety, the $k_t$ algorithm \cite{Catani:1993hr, Ellis:1993tq} exhibits distinct clustering behaviour that can significantly alter the logarithmic structure in two respects: by modifying the magnitude of NGLs, and by inducing a new, large tower of primary‑emission clustering logarithms (CLs).

In our previous work \cite{Ziani:2021dxr}, we computed the jet mass distribution for vector boson+jet (V+jet) and Higgs+jet (H+jet) processes at hadron colliders, providing complete two-loop results for NGLs in the anti‑$k_t$ algorithm and for NGLs and CLs in the $k_t$ and Cambridge/Aachen algorithms \cite{Dokshitzer:1997in, Wobisch:1998wt}. These calculations revealed several notable features, including the persistence of NGLs in the small‑$R$ limit (the “edge effect”), the dominance of gluon‑initiated jets owing to their larger colour factors, and the substantial reduction of NGLs when using the $k_t$ algorithm compared with anti‑$k_t$.
In a subsequent study \cite{Khelifa-Kerfa:2024udm}, we extended the fixed‑order NGLs calculations of \cite{Ziani:2021dxr} to four-loop order for the anti‑$k_t$ algorithm, demonstrating improved agreement with all‑orders numerical resummation and shedding new light on the convergence of the perturbative series. These results build upon the findings of \cite{Khelifa-Kerfa:2024roc}, which elucidated the structure of $k_t$ clustering to all orders in perturbation theory across various QCD processes.

Despite these advances, a comprehensive understanding of jet‑mass distributions with $k_t$ clustering beyond two-loops has remained elusive. The $k_t$ algorithm introduces further complexity through its propensity to cluster softer emissions first, which may either suppress or enhance logarithmic contributions depending on the emission topology (primary versus secondary gluons). This clustering behaviour gives rise to clustering logarithms (CLs), which first appear at two-loops and exhibit non‑trivial dependence on the jet radius $R$ and on colour factors \cite{Appleby:2002ke, Banfi:2005gj, Delenda:2006nf, Khelifa-Kerfa:2011quw, Delenda:2012mm, Bouaziz:2022tik, Ziani:2021dxr, Benslama:2023gys, Becher:2023znt, Khelifa-Kerfa:2024roc, Khelifa-Kerfa:2025cdn}. Recent studies of $e^+e^-$ annihilation processes \cite{Khelifa-Kerfa:2024hwx, Khelifa-Kerfa:2024gyv} have computed CLs up to four and six loops, confirming the characteristic features observed at lower orders—most notably the “edge effect” analogous to that of NGLs.

This paper extends the work of \cite{Khelifa-Kerfa:2024udm} to the case of $k_t$ clustering for V/H+jet processes at hadron colliders. We compute the full distribution of the invariant mass‑squared of the leading jet, including both CLs and NGLs, up to four-loops in perturbation theory. Our calculations are performed within the eikonal approximation under the assumption of strong energy ordering, achieving single‑logarithmic accuracy. In particular, we provide the first four-loop calculation of the jet‑mass distribution with $k_t$ clustering for all relevant partonic channels in V/H+jet production and derive semi‑analytical expressions that retain the complete dependence on colour factors and the jet radius $R$, thus enabling detailed studies of the colour- and $R$‑dependence of the logarithmic coefficients.

We have systematically compared the $k_t$ results with their anti‑$k_t$ counterparts, quantified the reduction of NGLs and identified regions in which $k_t$ clustering may unexpectedly enhance logarithmic contributions. In particular, for certain partonic channels and over a range of jet radii, we observe that NGLs in $k_t$ clustering can be of the same magnitude as—or even larger than—their anti‑$k_t$ equivalents. This behaviour first arises at three-loops and was therefore not apparent in our previous two‑loop calculations \cite{Ziani:2021dxr}. Moreover, at four-loops we find that the coefficients of CLs may change sign for sufficiently large values of the jet radius in specific channels. Both of these features are absent in $e^+e^-$ processes and have not been reported previously.

Furthermore, we have compared our fixed-order results with the output of the numerical Monte Carlo (MC) resummation code of ref. \cite{Dasgupta:2001sh}, demonstrating improved agreement for most partonic channels when higher-loop contributions are included, for both NGLs and CLs. We have also assessed the impact of the said higher loop orders on the full resummed form factor of the jet mass observable in the specific process of Z+jet. The reduction in the peak region of the distribution is found to be less than $7\%$ at two-loops and progressively smaller at three and four-loop orders.

The paper is organised as follows. In Section~\ref{sec:Kinematics} we introduce our notation, kinematic variables and the $k_t$ jet algorithm. Section~\ref{sec:FO} describes the fixed‑order calculation framework and presents our two, three and four-loop results for CLs and NGLs. Section~\ref{sec:Comparisons} offers a detailed comparison of fixed‑order predictions with all‑orders resummation, and Section~\ref{sec:Conclusion} summarises our conclusions and outlines prospects for future work.

\section{Definitions}
\label{sec:Kinematics}

\subsection{Kinematics}

Consider the production of a single jet in association with a Higgs (H) or vector (V = Z, $W^{\pm}$, $\gamma$) boson at a hadron collider. For vector‑boson processes, there are three partonic Born channels to consider: $q\qb \to gV$, $qg \to qV$ and $\qb g \to \bar q V$. The latter two channels are equivalent, so we shall retain only the former, namely $qg \to qV$. In the case of Higgs production, there are four Born channels; three coincide with those of V+jet production, and the fourth is $gg \to gH$. Accordingly, we treat three Born channels:
\begin{align}\label{eq:Def:BornChannels}
(\delta_1)\colon\; q\bar q \to g\,V/H, \qquad
(\delta_2)\colon\; q\,g \to q\,V/H, \qquad
(\delta_3)\colon\; g\,g \to g\,H.
\end{align}
From the perspective of our QCD calculations, all three channels are identical, as each involves three hard coloured partons and a colour‑neutral boson. The precise nature of the boson does not affect the QCD dynamics, nor does the flavour of the $W^{\pm}$‑mediated processes, since all quark flavours are treated on an equal footing. The Born channels differ only in their (a) Born cross sections and (b) associated colour factors.

State‑of‑the‑art calculations of the total cross sections for these processes have reached next‑to‑next‑to‑next‑to‑leading order (N$^3$LO) accuracy \cite{Anastasiou:2016cez, Mistlberger:2018etf, Chen:2019fhs, Chen:2025utl}, with next‑to‑next‑to‑leading order (NNLO) results available for some time \cite{Gauld:2021ule, Boughezal:2015aha, Boughezal:2015ded, Caola:2015wna, Gehrmann-DeRidder:2015wbt, Boughezal:2016dtm, Gehrmann-DeRidder:2016cdi, Campbell:2017dqk}. In recent years, dedicated parton‑level event generators such as \texttt{NNLOJET} \cite{NNLOJET:2025rno} have been developed to automate the calculation of QCD jet cross sections at NNLO accuracy.

For the partonic channels defined in \eqref{eq:Def:BornChannels}, we parametrise the four‑momenta of the hard coloured partons and of subsequently emitted soft gluons as
\begin{align}\label{eq:Def:Momenta}
p_a &= x_a \,\frac{\sqrt{s}}{2}\,(1,0,0,1), \notag\\
p_b &= x_b \,\frac{\sqrt{s}}{2}\,(1,0,0,-1), \notag\\
p_j &= p_t\,(\cosh y,\cos\varphi,\sin\varphi,\sinh y), \notag\\
k_i &= k_{ti}\,(\cosh\eta_i,\cos\phi_i,\sin\phi_i,\sinh\eta_i),
\end{align}
where $a$ and $b$ label the incoming partons and $j$ labels the outgoing hard parton, which has transverse momentum $p_t$, rapidity $y$ and azimuthal angle $\varphi$ with respect to the beam axis (taken to be along the $z$‑axis). The incoming partons carry momentum fractions $x_a$ and $x_b$ of the partonic centre‑of‑mass energy $\sqrt{s}$. The soft gluons $g_i$ have transverse momenta $k_{ti}$, rapidities $\eta_i$ and azimuthal angles $\phi_i$. The distributions of the parton momentum fractions are described by the standard parton distribution functions (PDFs), which are now available up to N$^3$LO accuracy \cite{Barontini:2024eii}. Recoil effects enter at next‑to‑next‑to‑leading logarithmic (NNLL) accuracy and beyond \cite{Banfi:2004yd} and are henceforth neglected, as they are beyond the scope of the current work. All partons are treated as massless; extensions to incorporate heavy‑flavour quarks will be presented elsewhere.

\subsection{Observable and jet algorithm}
\label{subsec:Observable+jetAlgo}

We study the typical non‑global QCD observable, namely the invariant mass‑squared of the leading hard jet normalised to the square of its transverse momentum \cite{Dasgupta:2012hg, Ziani:2021dxr, Khelifa-Kerfa:2024udm},
\begin{subequations}
\begin{align}\label{eq:Def:JetMass}
 \varrho \;=\;\frac{m_j^2}{p_t^2}
              \;=\;\frac{1}{p_t^2}\left(p_j+\sum_{i\in j}k_i\right)^2
              \;=\;\sum_{i\in j}\varrho_i+\cO\!\left(k_{t}^2/p_t^2\right),
\end{align}
where the sum runs over all soft gluons that are clustered {\it inside} the jet by the chosen algorithm, while emissions outside the jet do not contribute to its mass. In the soft (eikonal) limit we neglect terms scaling as $k_t^2/p_t^2$. The contribution of a single soft gluon $k_i$ to the normalised jet‑mass, $\varrho_i$, is given by
\begin{align}\label{eq:Def:JetMass-SingleGluon}
 \varrho_i
   = 2\,\frac{p_j\cdot k_i}{p_t^2}
   = 2\,\xi_i\left[\cosh(\eta_i-y)-\cos(\phi_i-\varphi)\right],
\end{align}
\end{subequations}
with the momentum fraction $\xi_i\equiv k_{ti}/p_t$.
It is convenient to reparametrise the angular separations in terms of polar variables $(r_i,\theta_i)$ \cite{Ziani:2021dxr, Khelifa-Kerfa:2024udm},
\begin{align}\label{eq:Def:PolarParametrisation}
 \eta_i-y = R\,r_i\cos\theta_i,
 \qquad
 \phi_i-\varphi = R\,r_i\sin\theta_i,
\end{align}
where $r_i>0$, $0<\theta_i<2\pi$ and $R$ is the jet‑radius parameter. Expanding the single‑gluon contribution \eqref{eq:Def:JetMass-SingleGluon} in powers of $R$ yields
\begin{align}\label{eq:Def:JetMass-Expanded}
 \varrho_i
   = \xi_i\left[R^2\,r_i^2
     + \tfrac{1}{12}R^4\,r_i^4\cos(2\theta_i)
     + \cdots\right],
\end{align}
where the ellipsis denotes terms of order $R^6$ and beyond. Since terms of order $R^4$ and higher can be shown to contribute only beyond single‑logarithmic accuracy, we retain solely the leading piece,
\begin{align}
 \varrho_i \simeq R^2\,\xi_i\,r_i.
\end{align}

The longitudinally invariant $k_t$ jet algorithm \cite{Catani:1993hr, Ellis:1993tq}, a member of the generalised sequential recombination family \cite{Cacciari:2011ma}, is defined as follows:

\begin{enumerate}
  \item Begin with an initial list of final‑state particles. For each pair $(i,j)$, compute the inter‑particle and beam distances
    \begin{align}\label{eq:Def:KTDistances}
      d_{ij} &= \min\left(p_{ti}^2,\,p_{tj}^2\right)\,\frac{\Delta R_{ij}^2}{R^2},
      & d_{iB} &= p_{ti}^2,
    \end{align}
    where
    \begin{align}
        \Delta R_{ij}^2 = (\eta_i-\eta_j)^2 + (\phi_i-\phi_j)^2,
    \end{align}
    with $\eta_i,\phi_i$ and $p_{ti}$ denoting the rapidity, azimuth and transverse momentum of particle $i$ relative to the beam axis.

  \item Identify the smallest distance among all $\{d_{ij},d_{iB}\}$. If this minimum is a $d_{ij}$, merge particles $i$ and $j$ into a single pseudo‑jet by summing their four‑momenta in the E‑scheme recombination. If the minimum is a $d_{iB}$, declare particle $i$ a final jet and remove it from the list.

  \item Repeat steps 1 and 2 until no particles remain to be clustered.
\end{enumerate}

The factor $\min(p_{ti}^2,p_{tj}^2)$ in \eqref{eq:Def:KTDistances} ensures that softer particles are clustered first. In particular, a softer particle $j$ will be clustered with particle $i$ (rather than being declared a jet) if
\begin{align}\label{eq:Def:ClusteringCond}
  \Delta R_{ij}^2 < R^2.
\end{align}
Under the polar parametrisation of eq.~\eqref{eq:Def:PolarParametrisation}, one finds  $
  \Delta R_{ij}^2 = R^2\left(r_i^2 + r_j^2 - 2r_ir_j\cos(\theta_i-\theta_j)\right)$, so that the clustering condition \eqref{eq:Def:ClusteringCond} becomes
\begin{align}\label{eq:Def:ClusteringCond-Polar}
  r_i^2 + r_j^2 - 2r_ir_j\cos(\theta_i-\theta_j) < 1.
\end{align}
In the strong‑ordering limit, emissions are ordered by softness, greatly simplifying the clustering sequence. Moreover, upon merging, the pseudo‑jet’s four‑momentum essentially coincides with that of the harder constituent.

The effect of $k_t$ clustering on the jet‑mass observable can be illustrated by considering a simple final‑state configuration of three particles: a jet‑initiating parton $p_j$, a harder soft gluon $k_h$, and a softer gluon $k_s$. Suppose $k_s$ is emitted within a distance $R$ of $p_j$, while $k_h$ lies outside this radius. In the absence of clustering (or when using the anti‑$k_t$ algorithm \cite{Cacciari:2008gp}), $k_s$ remains part of the jet initiated by $p_j$ and contributes to its mass. However, under $k_t$ clustering, if the separation between $k_s$ and $k_h$ is smaller than that between $k_s$ and $p_j$, the two gluons will recombine first. Consequently, $k_s$ may be “dragged out” of the jet by $k_h$ and no longer contribute to the jet mass. Conversely, if $k_h$ lies within the jet radius and $k_s$ initially lies outside, $k_t$ clustering can “drag in” $k_s$ towards $p_j$ by first recombining it with $k_h$, thereby causing $k_s$ to contribute to the mass. These drag‑in and drag‑out mechanisms operate equally for any number of soft emissions and apply to both primary and secondary gluon radiation.

The foregoing considerations, corroborated by earlier studies (see, for example, \cite{Appleby:2002ke, Banfi:2005gj, Delenda:2006nf, Banfi:2010pa, Khelifa-Kerfa:2011quw, Delenda:2012mm, Becher:2023znt, Khelifa-Kerfa:2024udm, Khelifa-Kerfa:2024hwx}), demonstrate that $k_t$ clustering can convert gluon configurations which would not contribute to the jet mass in the absence of clustering into configurations that do. This mechanism underlies the emergence of CLs. Furthermore, clustering introduces additional phase‑space constraints on the remaining contributing (secondary correlated emissions) configurations, thereby reducing the available phase space and, in particular, suppressing the coefficients of NGLs.

\subsection{Observable distribution}
\label{subsec:ObservableDist}

At next‑to‑leading logarithmic (NLL) accuracy, the differential (jet‑shape) cross‑section for our non‑global observable, the normalised invariant jet mass, in a specific channel $\delta$, may be written as \cite{Dasgupta:2012hg, Ziani:2021dxr, Khelifa-Kerfa:2024udm}
\begin{align}\label{eq:Def:DiffJetMassXsection}
 \frac{\d\S_\delta(\rho)}{\d\cB_\delta}
 = \int_0^\rho \frac{\d^2\s_\delta}{\d\cB_\delta\,\d\varrho}\,\d\varrho,
\end{align}
where $\rho$ denotes the jet‑mass veto, $\d\cB_\delta$ is the differential element of the Born configuration for channel $\delta$, and $\d\s_\delta$ is the partonic differential cross‑section for that channel. The integrated jet‑mass distribution is then obtained by imposing a set of Born‑level kinematic cuts, $\Xi_{\cB}$, and summing over all partonic channels:
\begin{align}\label{eq:Def:IntegJetMassXsection}
 \S(\rho)
 = \sum_{\delta}\int \d\cB_\delta \;\frac{\d\S_\delta(\rho)}{\d\cB_\delta}\;\Xi_{\cB}.
\end{align}
In this paper, we concentrate solely on the perturbative calculation of the CLs and NGLs appearing in the differential distribution \eqref{eq:Def:DiffJetMassXsection} up to four-loops. The reader is referred to ref.\ \cite{Ziani:2021dxr} for details of the integration, including the treatment of parton distribution functions and scale choices.

In the phase‑space region where the jet mass is small, $\rho \ll 1$, the perturbative distribution \eqref{eq:Def:DiffJetMassXsection} is dominated by large logarithms and may be written as
\begin{align}\label{eq:Def:DiffJetMass-PT}
  \frac{\d\S_\delta(\rho)}{\d\cB_\delta}
  &= \frac{\d\sigma_{0,\delta}}{\d\cB_\delta}\,
     f_{\cB,\delta}(\rho)\,\left[1 + \cO(\alpha_s)\right],
\end{align}
where $\d\sigma_{0,\delta}/\d\cB_\delta$ is the partonic Born differential cross‑section for channel $\delta$, and $\cO(\alpha_s)$ denotes non‑logarithmic corrections suppressed by higher powers of the strong coupling, $\alpha_s$. The function $f_{\cB,\delta}(\rho)$ resums the large logarithms of the jet‑mass observable and, for the $k_t$ algorithm, factorises as \cite{Delenda:2006nf, Delenda:2012mm, Khelifa-Kerfa:2024roc}
\begin{align}\label{eq:Def:ResummedJetMass}
  f_{\cB,\delta}(\rho)
  = f_{\cB,\delta}^{\rm global}(\rho)\,
    \cS_\delta(\rho)\,
    \cC_\delta(\rho),
\end{align}
where $f_{\cB,\delta}^{\rm global}(\rho)$ resums logarithms arising from soft‑collinear, hard‑collinear and soft-wide-angle primary emissions off all incoming and outgoing hard partons. This global factor is algorithm-independent and was computed in detail for the same observable in refs.\ \cite{Dasgupta:2012hg, Ziani:2021dxr}; it will not be repeated here. The reader is referred to those works (and their appendices) for further details.

In the present paper we focus on the jet‑algorithm dependent functions $\cS_\delta(\rho)$ and $\cC_\delta(\rho)$, which encode the resummation of NGLs and CLs, respectively. Unlike the global resummed factor, no all‑orders {\it analytical} expressions for these functions are known, so a fixed‑order perturbative treatment is indispensable for an analytic investigation of their structure. Accordingly, we expand
\begin{align}\label{eq:Def:JetMass-FOExpansion}
 f_{\cB,\delta}(\rho)
 &= 1 + f^{(1)}_{\cB,\delta}(\rho)
     + f^{(2)}_{\cB,\delta}(\rho)
     + \cdots,
\end{align}
where $f^{(n)}_{\cB,\delta}(\rho)$ denotes the $n$‑loop contribution in $\alpha_s$ to the jet‑mass distribution, comprising the NGLs and CLs components $\cS_{n,\delta}(\rho)$ and $\cC_{n,\delta}(\rho)$, respectively. In the following sections, we present detailed two, three and four-loop calculations of $f_{\cB,\delta}(\rho)$ for all three partonic channels defined in eq.~\eqref{eq:Def:BornChannels}.

\section{Fixed-order calculations}
\label{sec:FO}

Following the procedure of the measurement operator $\Uh$, first introduced in \cite{Schwartz:2014wha} and subsequently employed in our studies of NGLs and CLs \cite{Khelifa-Kerfa:2015mma, Ziani:2021dxr, Khelifa-Kerfa:2024roc, Khelifa-Kerfa:2024udm, Khelifa-Kerfa:2024gyv, Khelifa-Kerfa:2024hwx}, we express the $m$‑loop contribution to the partonic jet‑mass distribution as
\begin{align}\label{eq:FO:fB-mOrder}
 f_{\cB,\delta}^{(m)}(\rho)
 &= \sum_{\X} \int_{\xi_1>\xi_2>\cdots>\xi_m}
    \Biggl(\prod_{i=1}^m \d\Phi_i\Biggr)\,
    \Uh_m\,
    \cW_{1\ldots m,\delta}^\X\,
    \Xi_m^{k_t}(k_1,\ldots,k_m),
\end{align}
where $\cW_{1\ldots m,\delta}^\X$ denotes the eikonal amplitude squared for emission of $m$ strongly energy‑ordered soft gluons in configuration $X$ for channel $\delta$. Its general form for hadronic processes with three hard partons (such as V/H+jet processes) has been presented in \cite{Khelifa-Kerfa:2020nlc}, with explicit results up to four-loops.
Since each soft gluon may be either real (R) or virtual (V), a configuration \(X\) at \(m\)th order corresponds to one of the \(2^m\) real/virtual assignments of the \(m\) gluons. Consequently, the sum in equation \eqref{eq:FO:fB-mOrder} runs over all such configurations \(X\), each weighted by its specific eikonal amplitude squared \(\cW_{1\ldots m,\delta}^\X\).
The phase‑space element for gluon $i$ is
\begin{align}\label{eq:FO:PhaseSpaceFactor}
 \d\Phi_i
 &= \asb\,\frac{\d\xi_i}{\xi_i}\,\d\eta_i\,\frac{\d\phi_i}{2\pi}
  = \asb\,\frac{\d\xi_i}{\xi_i}\,R^2\,r_i\,\d r_i\,\d\theta_i,
\end{align}
where $\asb\equiv\as/\pi$ and the second equality follows from the parametrisation \eqref{eq:Def:PolarParametrisation}. The factor $\Xi_m^{k_t}(k_1,\ldots,k_m)$ is the clustering function which encodes the constraints imposed by the $k_t$ algorithm on the phase‑space of a given  configuration $X$ that contributes to the jet mass. Strong ordering in the momenta of the emitted soft gluons is enforced by the condition $\xi_1>\xi_2>\cdots>\xi_m$.

The measurement operator, at a given loop order $m$, factorises as \cite{Schwartz:2014wha, Khelifa-Kerfa:2015mma, Khelifa-Kerfa:2024roc}
\begin{align}\label{eq:FO:MeasOperator}
 \Uh_m = \prod_{i=1}^m \uh_i,
 \qquad
 \uh_i = 1 - \Theta_i^{\rm R}\,\Theta_i^\rho\,\Theta_i^{\rm in},
\end{align}
where $\Theta_i^{\rm R}=1$ for a real emission and zero otherwise,
\begin{align}
  \Theta_i^{\rm in}
 = \Theta\left[R^2 - (\eta_i-y)^2 - (\phi_i-\varphi)^2\right]
 = \Theta(1 - r_i^2),
\end{align}
and $\Theta_i^\rho=\Theta(\varrho_i - \rho)$ restricts the single-gluon contribution to the jet mass to be greater than the veto $\rho$. Emissions with $r_i^2>1$ satisfy $\Theta_i^{\rm out}=1-\Theta_i^{\rm in}=\Theta(r_i^2-1)$ and lie outside the jet.

Before proceeding to the new higher‑loop calculations, we briefly recall the one-loop result from ref.~\cite{Ziani:2021dxr}. Following the approach of refs.~\cite{Khelifa-Kerfa:2024roc, Khelifa-Kerfa:2024hwx}, the sum over the one-loop gluon configurations \(X = \mathrm{R},\mathrm{V}\) can be written as
\begin{align}\label{eq:1loop:SumX}
 \sum_\X \uh_1\,\cW_{1,\delta}^\X
 &= \uh_1\,\cW_{1,\delta}^\R + \uh_1\,\cW_{1,\delta}^\V
 = -\,\Theta_1^\rho\,\Theta_1^\inn\,\cW_{1,\delta}^\R\,.
\end{align}
Here \(\cW_{1,\delta}^\mathrm{R}\) and \(\cW_{1,\delta}^\mathrm{V}\) are the real and virtual one-loop eikonal amplitudes squared \cite{Khelifa-Kerfa:2020nlc}, given by
\begin{align}\label{eq:1loop:W1R-W1V}
 \cW_{1,\delta}^{\R} = \sum_{(i\ell) \in \Delta_\delta} \cC_{i\ell} w^1_{i\ell}, \qquad
 \cW_{1,\delta}^\V = - \cW_{1,\delta}^{\R},
\end{align}
with the dipole set \(\Delta_\delta = \{(ab), (aj), (bj)\}\) corresponding to the three hard partons \(p_a, p_b\) and \(p_j\) in channel \(\delta\). The associated colour factor is defined by
\begin{align}\label{eq:1loop:ColorFactorDefn}
 \cC_{i\ell} \equiv -2\,\mathbf{T}_i\!\cdot\!\mathbf{T}_\ell,
\end{align}
where \(\mathbf{T}_i\) are the generators of SU\((N_c)\).
For further details the reader is referred to refs.\ \cite{Khelifa-Kerfa:2020nlc} and \cite{Delenda:2015tbo}. The colour factors associated with the Born‑level dipoles in V/H+jet production are
\begin{align}\label{eq:1loop:ColorFactorsChannel}
 \cC_{q\bar q} = \cC_{qq} = 2\CF - \CA,
 \qquad
 \cC_{qg} = \cC_{gg} = \CA,
\end{align}
where the Casimir colour scalars are $\CF = (N_c^2-1)/(2N_c)$ and $\CA = N_c$. The one-loop dipole antenna function $w^i_{\alpha\beta}$ is defined by
\begin{align}\label{eq:1loop:AntennaFun}
 w_{\alpha\beta}^i = \frac{k_{ti}^2}{2}\,\frac{(p_\alpha\cdot p_\beta)}{(p_\alpha\cdot k_i)\,(k_i\cdot p_\beta)}.
\end{align}

Recall that at this order all jet algorithms are similar, so that \(\Xi_1(k_1)=\Theta_1^\mathrm{in}\). Substituting into the expression for the jet‑mass fraction \eqref{eq:FO:fB-mOrder} with \(m=1\), one obtains
\begin{align}\label{eq:1loop:fB-Integral}
 f_{\cB,\delta}^{(1)}(\rho) = - \int \d\Phi_1\,\Theta_1^\rho\,\cW_{1,\delta}^\mathrm{R}\,\Xi_1(k_1).
\end{align}
This integral has been evaluated in detail in refs.\ \cite{Dasgupta:2012hg, Ziani:2021dxr}, yielding, up to NLL accuracy,
\begin{align}\label{eq:1loop:fB-Final}
 f_{\cB,\delta}^{(1)}(\rho)
 = -\asb \left[\left(\cC_{aj}+\cC_{bj}\right)\frac{L^2}{4}
 + \left(\cC_{ab}\tfrac{R^2}{2} + \left(\cC_{aj}+\cC_{bj}\right)h(R)\right)L\right],
\end{align}
where \(L=\ln(R^2/\rho)\) and \(h(R)=R^2/8 + R^4/576 + \cO(R^8)\). Double‑logarithmic contributions arise exclusively from dipoles involving the outgoing parton \(p_j\), reflecting the soft and collinear singularities of the corresponding eikonal amplitude squared, whereas the incoming–incoming dipole \((ab)\) contributes only at single‑logarithmic order, since it lacks a collinear singularity. Exponentiation of the one-loop result \eqref{eq:1loop:fB-Final}, together with running‑coupling effects, produces the global (Sudakov) factor \(f_{\cB,\delta}^{\rm global}(\rho)\) in eq.~\eqref{eq:Def:ResummedJetMass}, which is universal and independent of the jet algorithm. Its full form is given in Refs.\ \cite{Dasgupta:2012hg, Ziani:2021dxr} and will not be repeated here.

In the following sections, we employ the results of ref.\ \cite{Khelifa-Kerfa:2024roc} to compute the fixed‑order expansion of \(f_{\cB,\delta}(\rho)\) up to four-loops. This work illustrates the application of \cite{Khelifa-Kerfa:2024roc} to hadronic collisions and extends the results of \cite{Khelifa-Kerfa:2024udm} to the case of \(k_t\) clustering.

\subsection{Two-loops}
\label{sec:2loop}

For the emission of two soft, strongly ordered gluons, \(k_1\) and \(k_2\), off a given partonic channel \(\delta\), the sum over all possible gluon real/virtual configurations of the eikonal amplitudes squared is \cite{Khelifa-Kerfa:2024roc, Khelifa-Kerfa:2024hwx}:
\begin{align}\label{eq:2loop:UWX}
 \sum_\X \Uh_2 \cW_{12, \delta}^{\X} = -\Theta_1^\rho \Theta_2^\rho\,\Theta_2^\inn \left[\cW_{12, \delta}^{\V\R} + \Theta_1^\out \Ob_{12}\, \cW_{12, \delta}^{\R\R}\right],
\end{align}
where \(\Ob_{i\ell}=1-\O_{i\ell}\) and
\(\O_{i\ell}=\Theta(d_{\ell B}-d_{i\ell})=\Theta(2r_i r_\ell\cos(\theta_i-\theta_\ell)-r_i^2)\),
with \(d_{\ell B}\) and \(d_{i\ell}\) defined in Sec.~\ref{subsec:Observable+jetAlgo} and the second equality following from the polar parametrisation \eqref{eq:Def:PolarParametrisation}.
The eikonal amplitudes squared at this order read \cite{Khelifa-Kerfa:2020nlc}:
\begin{align}\label{eq:2loop:EikAmpSq}
 &\cW_{12, \delta}^{\R\R} = \cW_{1,\delta}^\R\,\cW_{2,\delta}^\R + \oW_{12,\delta}^{\R\R}, \quad
 \cW_{12, \delta}^{\R\V} = -\cW_{12, \delta}^{\R\R}, \notag\\
 &\cW_{12, \delta}^{\V\R} = -\cW_{1,\delta}^\R\,\cW_{2,\delta}^\R, \quad
 \cW_{12, \delta}^{\V\V} = -\cW_{12, \delta}^{\V\R}.
\end{align}
The two‑loop irreducible contribution, \(\oW_{12,\delta}^{\R\R}\), is given by:
\begin{align}\label{eq:2loop:IrredEikAmp}
 \oW_{12,\delta}^{\R\R} = \CA \sum_{(i\ell) \in \Delta_\delta} \cC_{i\ell}\, \cA_{i\ell}^{12},
 \qquad
 \cA_{\alpha\beta}^{nm} = w_{\alpha\beta}^n\left(w_{\alpha n}^m + w_{n\beta}^m - w_{\alpha\beta}^m\right).
\end{align}
The function \(\cA_{\alpha\beta}^{nm}\) is known as the two‑loop antenna function.
Noting that \(\Theta_i^\inn + \Theta_i^\out =1\) and \(\O_{i\ell}+\Ob_{i\ell}=1\), eq. \eqref{eq:2loop:UWX} simplifies to:
\begin{align}\label{eq:2loop:UWX-B}
 \sum_\X \Uh_2 \cW_{12, \delta}^{\X} = -\Theta_1^\rho \Theta_2^\rho \left[
   -\Theta_1^\inn\Theta_2^\inn\,\cW_{1,\delta}^\R\cW_{2,\delta}^\R
   -\Theta_1^\out\Theta_2^\inn\,\O_{12}\,\cW_{1,\delta}^\R\cW_{2,\delta}^\R
   +\Theta_1^\out\Theta_2^\inn\,\Ob_{12}\,\oW_{12,\delta}^{\R\R}
   \right].
\end{align}
The first term in eq.~\eqref{eq:2loop:UWX-B} is independent of the clustering constraint \(\O_{i\ell}\) and therefore corresponds to the no‑clustering case; it is already accounted for by the expansion of the Sudakov form factor (the exponential of the one‑loop result \eqref{eq:1loop:fB-Final}). The second term in eq.~\eqref{eq:2loop:UWX-B} features the clustering constraint \(\O_{12}\) and involves only the primary‑emission component of the eikonal amplitude squared, thereby giving rise to the clustering logarithms at this order, \(\cC_{2,\delta}(\rho)\). The third term includes both the clustering constraint and the irreducible secondary‑emission contribution, and thus constitutes the non‑global logarithms at two-loops, \(\cS_{2,\delta}(\rho)\). From eq.~\eqref{eq:2loop:UWX-B}, one reads off the \(k_t\) clustering functions for CLs and NGLs:
\begin{align}\label{eq:2loop:ClusteringFuns}
 \Xi_{2, \text{\tiny cl}}^{k_t}(k_1,k_2) = \Theta_1^\out \Theta_2^\inn \O_{12}, \qquad
 \Xi_{2, \text{\tiny ng}}^{k_t}(k_1,k_2) = \Theta_1^\out \Theta_2^\inn \Ob_{12}.
\end{align}
Hence, at two-loops, the jet‑mass cross section assumes the form
\begin{align}\label{eq:2loop:fB}
 f_{\cB, \delta}^{(2)}(\rho) = \frac{1}{2!} \left[f_{\cB, \delta}^{(1)}(\rho)\right]^2 + \cC_{2,\delta}(\rho) + \cS_{2,\delta}(\rho).
\end{align}
Note that, according to eq.~\eqref{eq:2loop:ClusteringFuns}, CLs emerge only when the two gluons are clustered (through \(\O_{12}\)), whereas NGLs appear only when they remain unclustered (\(\Ob_{12}=1-\O_{12}\)).

\subsubsection{CLs}

\begin{figure}
\centering
\includegraphics[scale=0.51]{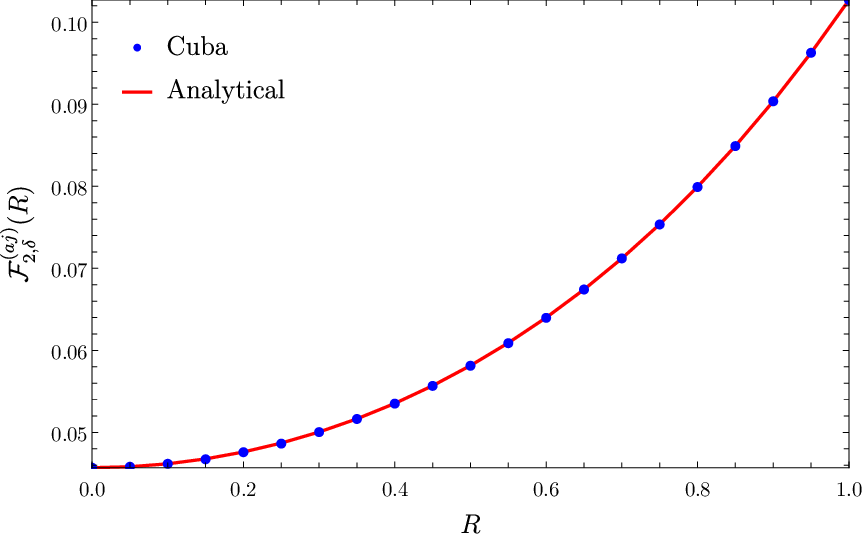}
\includegraphics[scale=0.51]{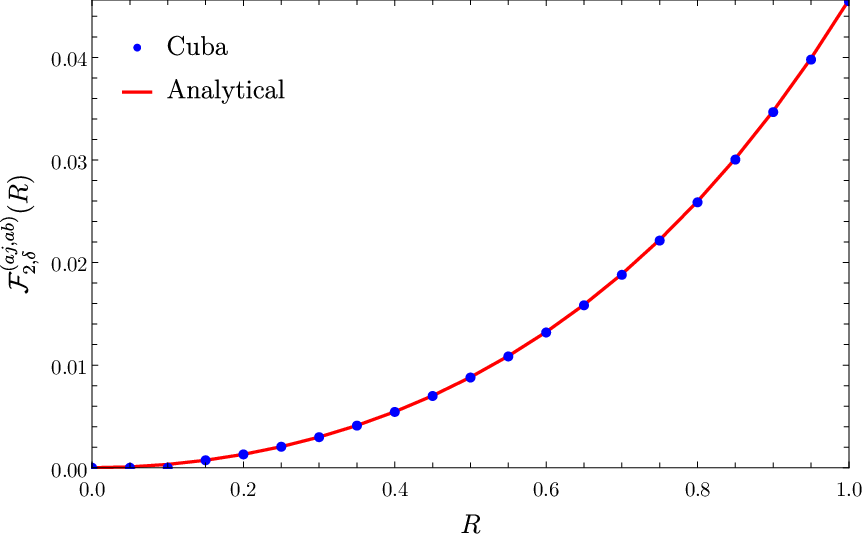}
\caption{Comparisons of the analytical and numerical results of some of the dipole contributions to the CLs coefficients at two-loops. }
\label{fig:2loop:F2-Cuba}
\end{figure}

Substituting the second term of eq.~\eqref{eq:2loop:UWX-B} into the master formula \eqref{eq:FO:fB-mOrder} and invoking eq.~\eqref{eq:1loop:W1R-W1V}, one obtains for the two‑loop clustering logarithms:
\begin{multline}\label{eq:2loop:CLs-A}
 \cC_{2, \delta}(\rho) = \sum_{(ik)\in\Delta_\delta}\sum_{(\ell m)\in\Delta_\delta}
 \cC_{ik}\,\cC_{\ell m}\,\asb^2\,R^4
 \int_{\xi_1>\xi_2}\prod_{i=1}^2 r_i\,\d r_i\,\frac{\d\theta_i}{2\pi}\,\frac{\d\xi_i}{\xi_i}\,
 \Theta\!\left(R^2r_i^2\xi_i-\rho\right)\,\times
\\
\Theta\!\left(r_1^2-1\right)\,\Theta\!\left(1-r_2^2\right)\,
\Theta\!\left(2r_1r_2\cos(\theta_1-\theta_2)-r_1^2\right)\,
w^1_{ik}\,w^2_{\ell m}\,.
\end{multline}
Since the one‑loop antenna functions \eqref{eq:1loop:AntennaFun} depend solely on angles, the energy integrals may be performed first. Working to NLL accuracy yields a factor of \(L^2/2!\,\Theta(r_2^2-\rho/R^2)\), where \(L=\ln(R^2/\rho)\). As the angular integrand remains finite for \(0<r_2<1\), one may set the lower limit of the \(r_2\) integral to zero without loss of accuracy at NLL. Accordingly, we recast the two‑loop CLs as
\begin{align}\label{eq:2loop:CLs-B}
 \cC_{2,\delta}(\rho) = \frac{1}{2!}\,\asb^2L^2\,\cF_{2,\delta}(R),
\end{align}
with the two‑loop CLs coefficient defined by
\begin{multline}\label{eq:2loop:F2-A}
 \cF_{2,\delta}(R) = \sum_{(ik)\in\Delta_\delta}\sum_{(\ell m)\in\Delta_\delta}
 \cC_{ik}\,\cC_{\ell m}\,R^4
 \int_1^\infty r_1\,\d r_1\,\frac{\d\theta_1}{2\pi}
 \int_0^1 r_2\,\d r_2\,\frac{\d\theta_2}{2\pi}\,\times
\\
\Theta\!\left(2r_2\cos(\theta_1-\theta_2)-r_1\right)\,
w^1_{ik}\,w^2_{\ell m}\,.
\end{multline}
The \(r_1\) integration is subject to \(r_1>1\), \(2r_2\cos(\theta_1-\theta_2)>r_1\) and \(\pi/(R\sin\theta_1)>r_1>-\pi/(R\sin\theta_1)\) (the latter ensuring \(\phi_1-\varphi\in[-\pi,\pi]\)), which together restrict \(r_1\in[1,2]\). The remaining integrals in eq.~\eqref{eq:2loop:F2-A} are evaluated by expanding the antenna functions in a power series in \(R\). Following ref.~\cite{Ziani:2021dxr}, one decomposes $\cF_{2,\delta}$ into a sum of two types of contributions, namely (independent) dipole and interference:
\begin{align}\label{eq:2loop:F2-B}
 \cF_{2,\delta}(R)
 = \sum_{(ik)\in\Delta_\delta}\cC_{ik}^2\,\cF_{2,\mathrm{dip}}^{(ik)}(R)
 + \sum_{\substack{(ik)\neq(\ell m)\\\in\Delta_\delta}}\cC_{ik}\,\cC_{\ell m}\,\cF_{2,\mathrm{int}}^{(ik,\ell m)}(R).
\end{align}
The first term in \eqref{eq:2loop:F2-B} corresponds to contributions from each of the three independent dipoles in channel \(\delta\), where both primary gluons are emitted sequentially from the same dipole, while the second term represents the interference between pairs of dipoles, with each gluon emitted from a different dipole. Carrying out the integrations, one obtains:
\begin{subequations}\label{eq:2loop:F2-dip+int}
\begin{align}
 \cF_{2, \text{dip}}^{(aj)}(R) &= \cF_{2, \text{dip}}^{(bj)}(R) = 0.0457 + 0.0475\,R^2 + 0.0091\,R^4 + 0.0004\,R^6 + \cO(R^8),
\label{eq:2loop:F2-dipA}
\\
 \cF_{2, \text{dip}}^{(ab)}(R) &= 0.052\,R^4,
\label{eq:2loop:F2-dipB}
\end{align}
and
\begin{align}
 \cF_{2, \text{int}}^{(aj, bj)}(R) &= \cF_{2, \text{int}}^{(bj, aj)}(R) = 0.0457 + 0.0042\,R^2 + 0.0004\,R^4 + 0.00004\,R^6 + \cO(R^8),
\\
 \cF_{2, \text{int}}^{(aj, ab)}(R) &= \cF_{2, \text{int}}^{(bj, ab)}(R) = 0.032\,R^2 + 0.013\,R^4 + 0.0006\,R^6 + \cO(R^8),
\\
 \cF_{2, \text{int}}^{(ab, aj)}(R) &= \cF_{2, \text{int}}^{(ab, bj)}(R) = 0.071\,R^2 + 0.013\,R^4 + 0.0003\,R^6 + \cO(R^8).
\end{align}
\end{subequations}
All analytical expressions have been cross‑checked against numerical integrations performed with the multidimensional \texttt{Cuba} library \cite{Hahn:2004fe}, using the \((\eta,\phi)\) parametrisation of eq.~\eqref{eq:Def:Momenta}, as illustrated in fig.~\ref{fig:2loop:F2-Cuba}.

\begin{figure}[t!]
\centering
\includegraphics[scale=0.7]{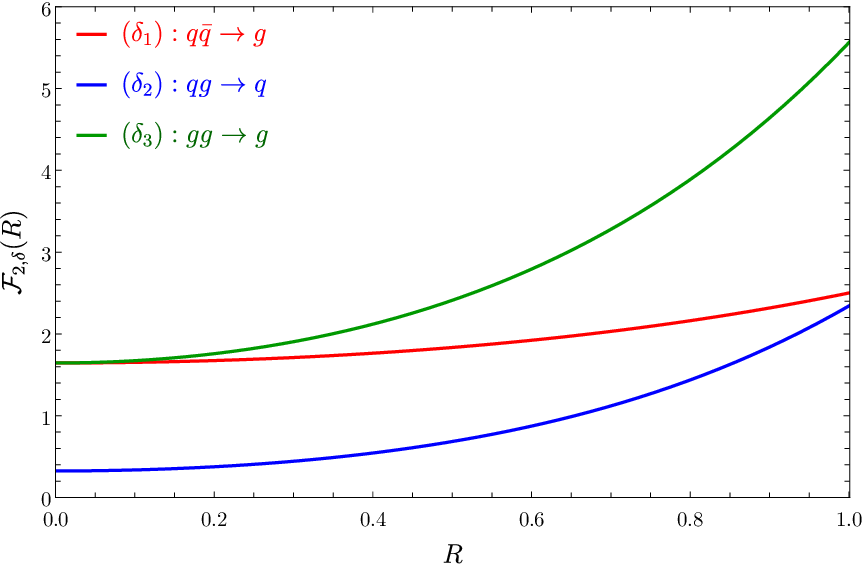}
\caption{The CLs coefficients at two-loops for each channel, for the $k_t$ jet algorithm.}
\label{fig:2loop:F2}
\end{figure}

Summing both dipole and interference contributions for each channel according to eq.~\eqref{eq:2loop:F2-B}, we obtain:
\begin{subequations}
\begin{align}
 \cF_{2, \delta_1}(R) &= \CFsq \left[0.208\,R^4\right] + \CF\CA \left[0.412\,R^2 -0.104\,R^4 +0.004\,R^6 \right] + \notag\\
&\quad+ \CAsq \left[0.183 - 0.019\,R^2 + 0.019\,R^4 - 0.001\,R^6\right] + \cO(R^8),
\end{align}
for channel ($\delta_1$),
\begin{align}
 \cF_{2, \delta_2}(R) &= \CFsq \left[0.183 + 0.190\,R^2 + 0.036\,R^4 + 0.002\,R^6\right]  +\CF\CA \Bigg[0.033\,R^2 + 0.017\, R^4 + \notag\\
&\quad+ 0.004\,R^6 \Bigg] + \CAsq \left[0.087\,R^2 + 0.069\, R^4 + 0.001\,R^6 \right] + \cO(R^8),
\end{align}
for channel ($\delta_2$), and
\begin{align}
 \cF_{2, \delta_3}(R) &= \CAsq \left[0.183 + 0.309\,R^2 + 0.123\,R^4 + 0.003\,R^6\right] + \cO(R^8),
\end{align}
for channel ($\delta_3$).
\end{subequations}
These coefficients are displayed in fig.~\ref{fig:2loop:F2}. As \(R\to0\), each coefficient approaches a non‑zero constant. From eqs.~\eqref{eq:2loop:F2-dip+int} one sees that this constant arises only when both gluons are emitted from dipoles containing the jet‑initiating parton \(p_j\). If one or both emitting dipoles do not involve \(p_j\), the constant term vanishes. Physically, this reflects the fact that for the \(k_t\) clustering condition \(\O_{12}\) to hold, the separation between the two gluons must be smaller than the distance of \(k_2\) to \(p_j\), which can occur only when both emissions originate from \(p_j\). A similar feature has been observed for CLs in \(e^+e^-\) annihilation \cite{Banfi:2010pa, Khelifa-Kerfa:2011quw, Delenda:2012mm, Kerfa:2012yae, Khelifa-Kerfa:2024hwx}, with the small‑\(R\) limit yielding the same numerical value \(0.183\, \text{C}_i^2\) (\(\text{C}_i^2=\CFsq\) for quark-initiated jets, \(\CAsq\) for gluon-initiated jets).
This behaviour can be traced to the fact that the eikonal amplitude squared for a soft‑gluon emission is maximally singular in the collinear limit. Consequently, whenever such an emission occurs—regardless of the jet‑radius \(R\)—there remains a finite contribution from the soft‑collinear region.

Moreover, fig.~\ref{fig:2loop:F2} demonstrates that gluon‑initiated jets exhibit larger CLs coefficients than quark-initiated jets. This is a direct consequence of the dependence on Casimir colour factors (cf.\ eq.~\eqref{eq:2loop:F2-dip+int}), with \(\CF=4/3\) and \(\CA=3\) in QCD. Naturally, channel ($\delta_3$) yields the largest contribution, since its coefficient scales purely as \(\CAsq\). We also observe an increase of the coefficients with the jet radius \(R\), reflecting the enlarged phase space for emissions within the jet region and the influence of initial‑state radiation. Compared to analogous results in \(e^+e^-\) annihilation \cite{Delenda:2012mm, Khelifa-Kerfa:2024hwx}, the CLs coefficients for V/H+jet processes are larger, owing to the appearance of mixed and purely gluonic colour channels (\(\CF\CA\) and \(\CAsq\)).

\subsubsection{NGLs}

\begin{figure}
\centering
\includegraphics[scale=0.51]{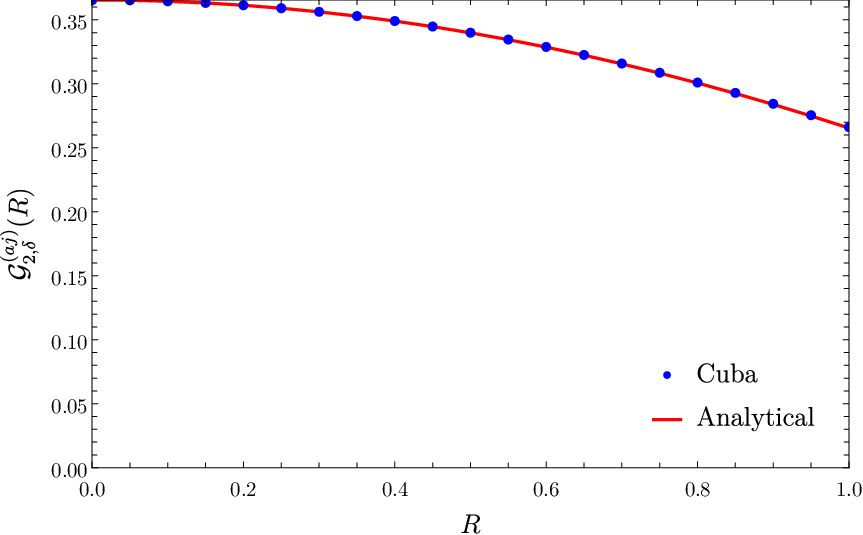}
\includegraphics[scale=0.51]{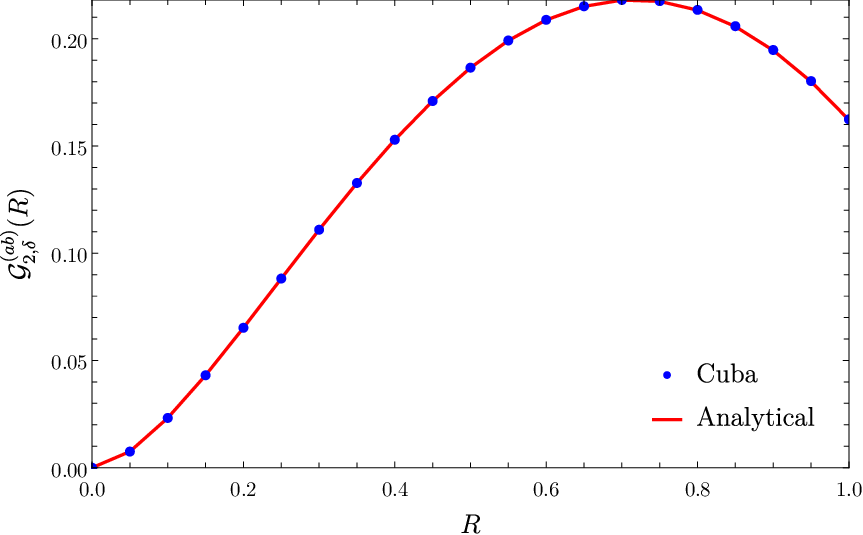}
\caption{Comparisons of the analytical and numerical results of some of the dipole contributions to the NGLs coefficients at two-loops. }
\label{fig:2loop:G2-Cuba}
\end{figure}

The third term in eq.~\eqref{eq:2loop:UWX-B} corresponds to correlated secondary emissions and hence to non‑global logarithms.  Substituting this term, together with the clustering function \eqref{eq:2loop:ClusteringFuns}, back into the general formula \eqref{eq:FO:fB-mOrder}, one finds for the two‑loop NGLs contribution in polar coordinates \eqref{eq:Def:PolarParametrisation}:
\begin{multline}\label{eq:2loop:NGLs-A}
 \cS_{2, \delta}(\rho) = -\CA \sum_{(ik) \in \Delta_\delta} \cC_{ik}\, \asb^2\,R^4
 \int_{\xi_1 > \xi_2} \prod_{i=1}^2 r_i \,\d r_i\,\frac{\d\theta_i}{2 \pi}\,\frac{\d\xi_i}{\xi_i}\,
 \Theta\!\left(R^2 r_i^2 \xi_i -\rho\right)\,\times  \\
 \times \Theta(r_1 -1)\,\Theta(1-r_2)\,\Theta\left(r_1 - 2 r_2 \cos(\theta_1-\theta_2)\right)\,\cA_{ik}^{12}.
\end{multline}
Proceeding as for the CLs, the \(\xi\)–integrals factorise and, to NLL accuracy, produce \(\tfrac12L^2\), so that
\begin{align}\label{eq:2loop:NGLs-B}
 \cS_{2,\delta}(\rho) = -\tfrac12\,\asb^2\,L^2\, \cG_{2,\delta}(R),
\end{align}
with the two‑loop NGLs coefficient defined by
\begin{subequations}\label{eq:2loop:G2-A}
\begin{align}
 \cG_{2, \delta}(R) &= \CA \sum_{(ik) \in \Delta_\delta} \cC_{ik}\, \cG_{2}^{(ik)}(R),
\label{eq:2loop:G2-channel}
\\
 \cG_2^{(ik)}(R) &= R^4
 \int_{1}^{\frac{\pi}{|R\sin\theta_1|}} r_1 \,\d r_1\,\frac{\d\theta_1}{\pi}
 \int_0^1 r_2 \,\d r_2\,\frac{\d\theta_2}{\pi}\,
 \Theta\left(r_1 - 2 r_2 \cos(\theta_1 - \theta_2)\right)\,\cA_{ik}^{12},
\end{align}
\end{subequations}
where the upper limit on $r_1$ comes from the fact that $\phi_1 - \varphi \in [-\pi,\pi]$, and recalling that $\sin\theta_1$ changes sign over the interval $[0,2\pi]$. To perform the above integrals semi-analytically as a power series in $R$, we first expand the two-loop antenna function $\cA_{ik}^{12}$ in $R$, and then split the $r_1$-integral into three regions: $1<r_1<2$, $2<r_1<\pi/R$, and $\pi/R<r_1<\pi/|R \sin\theta_1|$. In the first region, the limits of the $r_1$-integral are free from any $R$ dependence, and the integration can therefore be performed numerically to determine the coefficients of $R^n$. In the second and third regions, the clustering step function is automatically satisfied since $2 r_2 \cos(\theta_1 - \theta_2) \leq 2$, and can thus be set to one. Although the integrals in the second region depend explicitly on $R$, they can still be performed analytically. For the third region, we make the transformation $r_1 \to R r_1$, which effectively absorbs any $R$ dependence into the limits of integration. We obtain for the various dipoles:
\begin{subequations}\label{eq:2loop:G2-B}
\begin{align}
 \cG_{2}^{(aj)}(R) &= \cG_2^{(bj)}(R) = 0.366 - 0.104\,R^2 + 0.003\,R^4 + 0.0001\,R^6 + \cO(R^8),
\\
 \cG_2^{(ab)}(R) &= -R^2 \ln R + 0.015\,R^2 + 0.151\,R^4 - 0.004\,R^6 + \cO(R^8).
\end{align}
\end{subequations}
These semi‑analytic results agree with pure numerical integrations performed using the \texttt{Cuba} library in the \((\eta,\phi)\) parametrisation, as shown in fig.~\ref{fig:2loop:G2-Cuba}.

Summing the dipole contributions according to eq.~\eqref{eq:2loop:G2-channel}, we obtain for the three channels
\begin{subequations}\label{eq:2loop:G2-channel-Final}
\begin{multline}
 \cG_{2, \delta_1}(R) = \CF\CA \left[-2R^2 \ln R + 0.031\,R^2 + 0.301\,R^4 - 0.008\,R^6\right]
\\
\quad+ \CAsq \left[0.731 + R^2\ln R - 0.223\,R^2 - 0.144\,R^4 + 0.004\,R^6\right] + \cO(R^8),
\end{multline}
for channel ($\delta_1$),
\begin{multline}
 \cG_{2, \delta_2}(R) = \CF\CA \left[0.731 - 0.207\,R^2 + 0.007\,R^4 + 0.0003\,R^6\right]
\\
\quad+ \CAsq \left[- R^2\ln R + 0.015\,R^2 + 0.151\,R^4 - 0.004\,R^6\right] + \cO(R^8),
\end{multline}
for channel ($\delta_2$), and
\begin{align}
 \cG_{2, \delta_3}(R) = \CAsq \left[0.731 - R^2 \ln R - 0.192\,R^2 + 0.158\,R^4 - 0.003\,R^6\right] + \cO(R^8),
\end{align}
for channel ($\delta_3$).
\end{subequations}

\begin{figure}[t]
\centering
\includegraphics[scale=0.7]{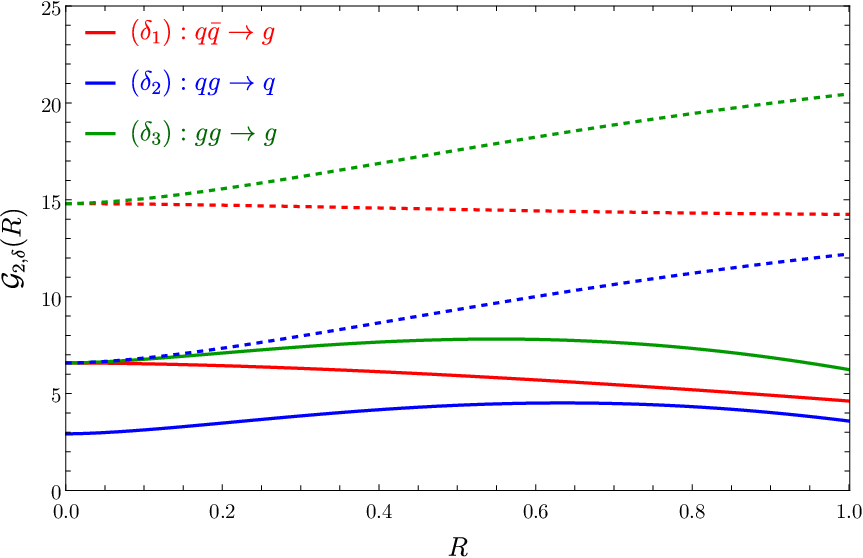}
\caption{The NGLs coefficients at two-loops for each channel for $k_t$ (solid lines) and ant-$k_t$ (dashed lines) jet algorithms.}
\label{fig:2loop:G2}
\end{figure}

In fig.~\ref{fig:2loop:G2} we display these two‑loop NGLs coefficients for all three channels. The edge (or boundary) effect is manifest in each case. That is, as \(R\to0\) the NGLs coefficients approach a finite constant rather than vanishing, consistent across both anti‑\(k_t\) (dashed curves) and \(k_t\) (solid curves) algorithms. Specifically,
\[
 \lim_{R \to 0} \cG_{2, \delta_1}(R) = \lim_{R \to 0} \cG_{2, \delta_3}(R) = 0.731\,\CAsq,
 \quad
 \lim_{R \to 0} \cG_{2, \delta_2}(R) = 0.731\,\CF\CA.
\]
The result for channel ($\delta_2$) precisely matches that for \(e^+e^-\) annihilation \cite{Banfi:2010pa, Khelifa-Kerfa:2011quw, Khelifa-Kerfa:2024hwx}. For channels ($\delta_1$) and ($\delta_3$), the small-\(R\) limits coincide exactly with those that would be obtained for gluon-initiated jets in the same \(e^+e^-\) context.

\begin{figure}[t]
\centering
\includegraphics[scale=0.7]{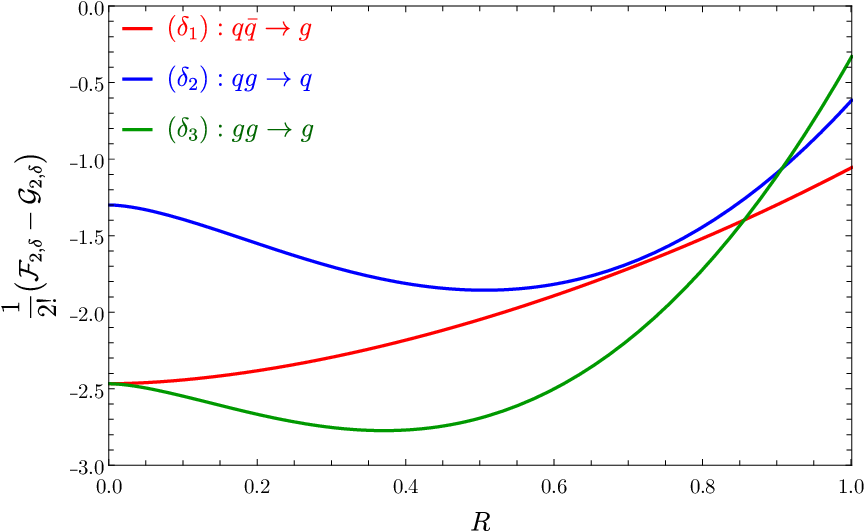}
\caption{The NGLs coefficients at two-loops for each channel for $k_t$ (solid lines) and ant-$k_t$ (dashed lines) jet algorithms.}
\label{fig:2loop:F2-G2}
\end{figure}

Moreover, the reduction of NGLs due to $k_t$ clustering, previously reported (see, for instance, \cite{Appleby:2002ke, Banfi:2005gj, Delenda:2006nf, Khelifa-Kerfa:2011quw}), is clearly visible in fig.~\ref{fig:2loop:G2}. While for the anti-$k_t$ algorithm each curve remains flat or increases with $R$ up to unity, the $k_t$ curves exhibit a turnover beyond \(R\gtrsim0.7\), particularly for channels ($\delta_2$) and ($\delta_3$). A comparable trend appears at the dipole level in fig.~\ref{fig:2loop:G2-Cuba}, notably for the incoming–incoming $(ab)$ dipole. This behaviour likely arises because, as the jet radius grows, the two soft emissions become increasingly prone to recombination, especially since for NGLs the softer gluon \(k_2\) is radiated off the harder gluon \(k_1\). Additional noteworthy features include the dominance of the gluon–gluon channel ($\delta_{3}$), reflecting its larger colour factor. To assess the combined impact of CLs and NGLs at this loop order, fig.~\ref{fig:2loop:F2-G2} displays the difference of their coefficients, \((\cC_{2,\delta} - \cG_{2,\delta})/2!\), from eqs.~\eqref{eq:2loop:CLs-B} and \eqref{eq:2loop:NGLs-B}. Although partially cancelling, these contributions remain substantial and cannot be neglected.

As noted above, the two-loop results in this section were first obtained in refs.\ \cite{Dasgupta:2012hg, Ziani:2021dxr}. To the best of our knowledge, no fixed-order perturbative calculations beyond two loops exist in the literature, and the principal objective of this paper is to present such results up to four-loop order.

\subsection{Three-loops}
\label{sec:3loop}

For the emission of three soft, $k_t$-ordered gluons from a partonic channel $(\delta)$, the sum over all real/virtual configurations is given by \cite{Khelifa-Kerfa:2024roc}:
\begin{multline}\label{eq:3loop:UWX-A}
 \sum_\X \Uh_3\, \cW_{123, \delta}^\X = - \Bigl(\prod_{i=1}^3 \Theta_i^\rho\Bigr)\,\Theta_3^\inn \Bigl[ \cW_{123,\delta}^{\V\V\R} + \Theta_1^\out \Ob_{13}\,\cW_{123,\delta}^{\R\V\R} + \Theta_2^\out \Ob_{23}\,\cW_{123,\delta}^{\V\R\R}\\
\quad+ \Theta_1^\out \bigl(\Theta_2^\out + \Theta_2^\inn \O_{12}\bigr)\,\Ob_{13}\,\Ob_{23}\,\cW_{123, \delta}^{\R\R\R}\Bigr],
\end{multline}
where the explicit forms of the eikonal amplitudes squared at this order are given in refs.\ \cite{Khelifa-Kerfa:2020nlc, Delenda:2015tbo}:
\begin{subequations}\label{eq:3loop:EikAmp}
\begin{align}
 \cW_{123,\delta}^{\R\R\R} &= \prod_{i=1}^3 \cW_{i,\delta}^\R + \sum_{j<k}\cW_{i,\delta}^\R\,\oW_{jk,\delta}^{\R\R} + \oW_{123,\delta}^{\R\R\R},
 &\cW_{123,\delta}^{\V\V\R} &= \prod_{i=1}^3 \cW_{i,\delta}^\R,\\
 \cW_{123,\delta}^{\V\R\R} &= -\prod_{i=1}^3 \cW_{i,\delta}^\R - \cW_{1,\delta}^\R\,\oW_{23,\delta}^{\R\R}, \\
 \cW_{123,\delta}^{\R\V\R} &= -\prod_{i=1}^3 \cW_{i,\delta}^\R - \sum_{k=2}^3\cW_{j,\delta}^\R\,\oW_{1k,\delta}^{\R\R} + \oW_{123,\delta}^{\R\V\R},
\end{align}
\end{subequations}
with $\cW_{i,\delta}^\R$ and $\oW_{ij,\delta}^{\R\R}$ defined in eqs.~\eqref{eq:1loop:W1R-W1V} and \eqref{eq:2loop:IrredEikAmp}, respectively.  The new three-loop irreducible contributions are
\begin{subequations}\label{eq:3loop:EikAmp-Irred}
\begin{align}
 \oW_{123,\delta}^{\R\R\R} &= \CAsq \sum_{(ij)\in\Delta_\delta}\cC_{ij}\bigl[\cA_{ij}^{12}\,\cAb_{ij}^{13} + \cB_{ij}^{123}\bigr] + \sum_{\pi_{\delta}} \cQ_\delta\bigl[\cG_{ij}^{k1}(2,3) + (2\leftrightarrow3)\bigr],\\
 \oW_{123,\delta}^{\R\V\R} &= -\CAsq \sum_{(ij)\in\Delta_\delta}\cC_{ij}\,\cA_{ij}^{12}\,\cAb_{ij}^{13} - \sum_{\pi_{\delta}} \cQ_\delta\bigl[\cG_{ij}^{k1}(2,3) + (2\leftrightarrow3)\bigr],
\end{align}
\end{subequations}
where \(\cAb_{ij}^{k\ell}=\cA_{ij}^{k\ell}/w_{ij}^k\), \(\pi_{\delta}=\{(ijk),(ikj),(jki)\}\), and the three-loop antenna and quadruple functions are defined by
\begin{align}\label{eq:3loop:B+G}
 \cB_{ij}^{k\ell m} &= w_{ij}^k\bigl(\cA_{ik}^{\ell m} + \cA_{jk}^{\ell m} - \cA_{ij}^{\ell m}\bigr),
 &\cG_{ij}^{k\ell}(m,n) &= w_{ij}^\ell\,T_{ij}^{k\ell}(n)\,U_{ij}^{k\ell}(m),
\end{align}
with the cross-channel functions
\begin{align}\label{eq:3loop:T+U}
 T_{ij}^{k\ell}(n) = w_{ij}^n + w_{k\ell}^n - w_{ik}^n - w_{j\ell}^n,\quad
 U_{ij}^{k\ell}(n) = w_{ij}^n + w_{k\ell}^n - w_{i\ell}^n - w_{jk}^n,
\end{align}
and the quadruple colour factors
\begin{align}\label{eq:3loop:Q_Col}
 \cQ_{\delta_1} = \cQ_{\delta_2} = \CAsq\bigl(\CA - 2\CF\bigr)=\CA,\qquad
 \cQ_{\delta_3} = 6\,\CA.
\end{align}
Using \(\Theta_i^\inn + \Theta_i^\out=1\), eq.~\eqref{eq:3loop:UWX-A} can be recast as
\begin{align}\label{eq:3loop:UWX-B}
 \sum_\X \Uh_3\, \cW_{123, \delta}^\X
 &= - \Bigl(\prod_{i=1}^3 \Theta_i^\rho\Bigr)\,\Theta_3^\inn \Bigl\{
    \Theta_1^\inn\Theta_2^\inn\,\cW_{123,\delta}^{\V\V\R}
    + \Theta_1^\inn\Theta_2^\out\bigl[\Ob_{23}\,\cW_{123,\delta}^{\V\R\R} + \cW_{123,\delta}^{\V\V\R}\bigr]\notag\\
 &\quad+ \Theta_1^\out\Theta_2^\inn\bigl[\O_{12}\Ob_{13}\Ob_{23}\,\cW_{123,\delta}^{\R\R\R} + \Ob_{13}\,\cW_{123,\delta}^{\R\V\R}
    + \cW_{123,\delta}^{\V\V\R}\bigr]\notag\\
 &\quad+ \Theta_1^\out\Theta_2^\out\bigl[\Ob_{13}\Ob_{23}\,\cW_{123,\delta}^{\R\R\R}
    + \Ob_{13}\,\cW_{123,\delta}^{\R\V\R}
    + \Ob_{23}\,\cW_{123,\delta}^{\V\R\R}
    + \cW_{123,\delta}^{\V\V\R}\bigr]
 \Bigr\}.
\end{align}

The term proportional to \(\Theta_1^\inn \Theta_2^\inn\) reproduces the no-clustering (anti-\(k_t\)) contribution already accounted for by the exponentiation of the one-loop result.  The contribution proportional to \(\Theta_1^\inn \Theta_2^\out\) may be recast using \(\O_{ij} + \Ob_{ij} = 1\):
\begin{align}
 \O_{23}\,\cW_{123,\delta}^{\V\V\R}
 + \Ob_{23}\,\bigl[\cW_{123,\delta}^{\V\R\R} + \cW_{123,\delta}^{\V\V\R}\bigr]
 = \cW_{1,\delta}^\R \times \O_{23}\,\cW_{2,\delta}^\R\,\cW_{3,\delta}^\R
 - \cW_{1,\delta}^{\R} \times \Ob_{12}\,\oW_{23,\delta}^{\R\R}.
\end{align}
These two terms correspond precisely to the interference of the one-loop piece \(f_{\cB,\delta}^{(1)}\) with the two-loop \((23)\) components of CLs and NGLs (\(\cC_{2,\delta}^{(23)}\) and  \(\cS_{2,\delta}^{(23)}\), respectively).  The term proportional to \(\Theta_1^\out \Theta_2^\inn\) contains both interference terms—namely products of \(f_{\cB,\delta}^{(1)}\) with the \((12)\) and \((13)\) components of \(\cC_{2,\delta}\) and \(\cS_{2,\delta}\)—and genuinely new irreducible contributions. Finally, the \(\Theta_1^\out \Theta_2^\out\) term yields purely three-loop irreducible clustering and non-global logarithms.  Consequently, the jet mass distribution at three loops takes the form:
\begin{align}\label{eq:3loop:fB}
 f_{\cB,\delta}^{(3)}(\rho) = \frac{1}{3!} \bigl[f_{\cB, \delta}^{(1)}\bigr]^2
 + f_{\cB, \delta}^{(1)} \times \bigl[\cC_{2,\delta} + \cS_{2,\delta}\bigr]
 + \cC_{3,\delta} + \cS_{3, \delta}.
\end{align}
We now proceed to discuss the irreducible three-loop contributions \(\cC_{3,\delta}\) and \(\cS_{3,\delta}\) in detail.

\subsubsection{CLs}

\begin{figure}[t!]
\centering
\includegraphics[scale=0.7]{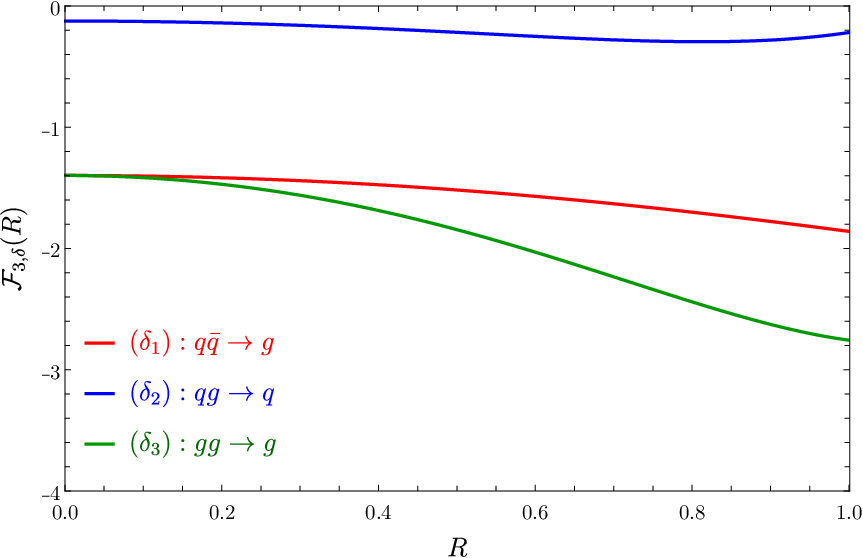}
\caption{The CLs coefficients at three-loops for each channel, for the $k_t$ jet algorithm.}
\label{fig:3loop:F3}
\end{figure}

The three-loop clustering logarithms can be shown to follow the same structural pattern as at two loops (cf.\ eq.~\eqref{eq:2loop:CLs-B}), namely
\begin{align}\label{eq:3loop:CLs}
 \cC_{3,\delta}(\rho) = -\frac{1}{3!}\,\asb^3\,L^3\,\cF_{3,\delta}(R),
\end{align}
where the CLs coefficient \(\cF_{3,\delta}(R)\) is given by \cite{Khelifa-Kerfa:2024hwx}
\begin{multline}\label{eq:3loop:F3-A}
 \cF_{3,\delta}(R) = \sum_{(ik)\in\Delta_\delta}\sum_{(\ell m)\in\Delta_\delta}\sum_{(ns)\in\Delta_\delta}
 \cC_{ik}\,\cC_{\ell m}\,\cC_{ns}\,R^6 \\[-6pt]
 \times \Bigg[\int_{1_\out}\int_{2_\out}\int_{3_\inn}\O_{13}\,\O_{23}
 + \int_{1_\out}\int_{2_\inn}\int_{3_\inn}\O_{12}\bigl(-1 + \Ob_{13}\Ob_{23}\bigr)\Bigg] \\
 \times \,w_{ik}^1\,w_{\ell m}^2\,w_{ns}^3,
\end{multline}
with the short-hand notations
\begin{align}\label{eq:3loop:ShortInteg}
 \int_{i_\out} \equiv \int_1^{\frac{\pi}{|R \sin\theta_i|}} r_i\,\d r_i
   \int_0^{2\pi}\frac{\d\theta_i}{2\pi},
 \quad
 \int_{i_\inn} \equiv \int_0^1 r_i\,\d r_i
   \int_0^{2\pi}\frac{\d\theta_i}{2\pi}.
\end{align}
As at two loops, one may decompose \(\cF_{3,\delta}(R)\) into dipole and interference contributions according to the emitting dipoles of each gluon:
\begin{multline}\label{eq:3loop:F3-B}
 \cF_{3,\delta}(R) = \sum_{(ik)\in\Delta_\delta} \cC_{ik}^3\,\cF_{2,\mathrm{dip}}^{(ik)}(R)
 + \sum_{\substack{(ik)\neq(\ell m)\\\in\Delta_\delta}} \cC_{ik}^2\,\cC_{\ell m}\,\cF_{2,\mathrm{dip-int}}^{(ik,\ell m)}(R)
 \\[-4pt]
 + \sum_{\substack{(ik)\neq(\ell m)\neq(ns)\\\in\Delta_\delta}} \cC_{ik}\,\cC_{\ell m}\,\cC_{ns}\,\cF_{2,\mathrm{int}}^{(ik,\ell m,ns)}(R).
\end{multline}
For each term in eq.~\eqref{eq:3loop:F3-B}, we evaluate the contributions from the three dipoles \((aj)\), \((bj)\) and \((ab)\), yielding \(3^3\) individual dipole combinations. The results of the numerical integrations are then fitted to a polynomial expansion in \(R\). In the interest of brevity, we present here only the combined results for each partonic channel:

\begin{subequations}\label{eq:3loop:F3-channels}
\begin{align}\label{eq:3loop:F2-d1}
 \cF_{3,\delta_1}(R) &= \CFcub \left[0.032\,R^6\right]
 + \CFsq \CA \left[0.001\,R^4 - 0.025\,R^6\right]\notag\\
 &\quad+ \CF \CAsq \left[-0.089\,R^2 + 0.0004\,R^4 + 0.009\,R^6\right]\notag\\
 &\quad+ \CAcub \left[-0.052 + 0.022\,R^2 - 0.0005\,R^4 - 0.0013\,R^6\right] + \cO(R^8),
\\
\label{eq:3loop:F2-d2}
 \cF_{3,\delta_2}(R) &= \CFcub \left[-0.052 - 0.053\,R^2 + 0.002\,R^4 + 0.004\,R^6\right]\notag\\
 &\quad+ \CFsq \CA \left[0.0007 + 0.015\,R^2 - 0.00009\,R^4 + 0.0016\,R^6\right]\notag\\
 &\quad+ \CF \CAsq \left[-0.0004 - 0.030\,R^2 - 0.005\,R^4 + 0.001\,R^6\right]\notag\\
 &\quad+ \CAcub \left[0.0003\,R^2 + 0.004\,R^4 + 0.008\,R^6\right] + \cO(R^8),
\\
\label{eq:3loop:F2-d3}
 \cF_{3,\delta_3}(R) &= \CAcub \left[-0.052 - 0.067\,R^2 + 0.001\,R^4 + 0.015\,R^6\right] + \cO(R^8).
\end{align}
\end{subequations}
The above results are depicted in fig.~\ref{fig:3loop:F3}.
All of the observations made at two loops persist at three loops. In particular, the boundary effect—where the CLs coefficients approach a constant as the jet radius \(R\) tends to zero—is again manifest. In this limit, one finds
\begin{align}
 &\lim_{R \to 0} \cF_{3,\delta_1} =  \lim_{R \to 0} \cF_{3,\delta_3} = -0.052\, \CAcub, \notag
\\
 &\lim_{R \to 0} \cF_{3,\delta_2} = -0.052\, \CFcub + 0.0007\,\CFsq\CA -0.0004\,\CF\CAsq.
\end{align}
These constants agree with those obtained for \(e^+e^-\) annihilation processes \cite{Delenda:2012mm, Khelifa-Kerfa:2024hwx}, except that the mixed \(\CFsq\CA\) and \(\CF\CAsq\) terms in channel \((\delta_2)\) have no analogue in the pure \(e^+e^-\) case. Channel \((\delta_3)\) (\(gg\to g\)) dominates numerically owing to its larger colour factor. Moreover, across the full range of \(R\), the magnitude of the three-loop coefficients $\cF_{3,\delta}/3!$ is smaller than at two loops for all three channels, indicating improved convergence of the perturbative series. Fig.~\ref{fig:3loop:F3} shows only a mild dependence of \(\cF_{3,\delta}\) on \(R\) for channels \((\delta_1)\) and \((\delta_2)\), and a more pronounced variation for channel \((\delta_3)\). From eq.~\eqref{eq:3loop:CLs} and fig.~\ref{fig:3loop:F3}, we note that the overall three-loop CLs contribution to the jet mass cross section is positive, as at two loops.

\subsubsection{NGLs}

Substituting the explicit formulae \eqref{eq:3loop:EikAmp} into \eqref{eq:3loop:UWX-B}, and considering only the terms contributing to NGLs while ignoring the overall factor in front of the curly parentheses in \eqref{eq:3loop:UWX-B} for now, we obtain:
\begin{align}\label{eq:3loop:UWX-NGLs_only}
+\Theta_1^\out \Theta_2^\inn &\Bigg[
\O_{12} \Ob_{13} \Ob_{23}\, \cW_{1,\delta}^\R \oW_{23,\delta}^{\R\R} + \Ob_{13} \left(\O_{12} \Ob_{23} - 1\right) \left[\cW_{2,\delta}^\R \oW_{13,\delta}^{\R\R} + \cW_{3,\delta}^\R \oW_{12,\delta}^{\R\R} \right] \\
&+ \O_{12} \Ob_{13} \Ob_{23} \left[\oW_{123,\delta}^{\R\R\R} +  \oW_{123,\delta}^{\R\V\R} \right]
+ \Ob_{13} (1-\O_{12} \Ob_{23})  \oW_{123,\delta}^{\R\V\R}
\Bigg] \\
+\Theta_1^\out \Theta_2^\out &\Bigg[
-\O_{13} \Ob_{23} \cW_{1,\delta}^\R \oW_{23,\delta}^{\R\R} - \Ob_{13} \O_{23} \left(\cW_{2,\delta}^\R \oW_{13,\delta}^{\R\R} + \cW_{3,\delta}^\R \oW_{12,\delta}^{\R\R} \right) \\
& + \Ob_{13} \Ob_{23} \left[\oW_{123,\delta}^{\R\R\R} + \oW_{123,\delta}^{\R\V\R} \right]
+ \Ob_{13} \O_{23} \oW_{123,\delta}^{\R\V\R}
\Bigg].
\end{align}
Note that for the part proportional to $\Theta_1^\out \Theta_2^\inn$, there are two interference terms in the first line: $\Ob_{13} \cW_{2, \delta}^{\R} \oW_{13,\delta}^{\R\R}$ and $\Ob_{12} \cW_{3,\delta}^\R\,\oW_{12,\delta}^{\R\R}$. The latter term can be identified through the simplification:
\begin{align}
 \Ob_{13} (\O_{12} \Ob_{23} -1) = -\Ob_{12} + \O_{13} - \O_{12} \left(1 - \Ob_{13} \Ob_{23}\right).
\end{align}
Apart from these two terms, the remainder of \eqref{eq:3loop:UWX-NGLs_only} represents new three-loop contributions. When the $k_t$ clustering is deactivated by setting all $\O_{ik} = 0$, eq.  \eqref{eq:3loop:UWX-NGLs_only} reduces to:
\begin{align}
 \Theta_1^\out \left(\Theta_2^\inn \, \oW_{123,\delta}^{\R\V\R} + \Theta_2^\out  \left[\oW_{123,\delta}^{\R\R\R} + \oW_{123,\delta}^{\R\V\R} \right] \right),
\end{align}
which matches the anti-$k_t$ formula for NGLs at three loops reported in \cite{Khelifa-Kerfa:2024udm} (eq. (3.15)). Analogous to the two-loop result in \eqref{eq:2loop:NGLs-B}, the three-loop NGLs contribution to the jet mass cross-section takes the form:
\begin{align}\label{eq:3loop:NGLs}
 \cS_{3,\delta}(\rho) = +\frac{1}{3!}\,\asb^3 \,L^3\, \left[\cG_{3,\mathrm{int}}^{\delta}(R) + \cG_{3, \mathrm{dip}}^{\delta}(R)  + \cG_{3, \mathrm{quad}}^{\delta}(R) \right],
\end{align}
where the three-loop NGLs coefficients (with the overall minus sign from \eqref{eq:3loop:UWX-B} absorbed into their definitions) are:
\begin{subequations}
\begin{align}\label{eq:3loop:G3Int}
\cG_{3,\mathrm{int}}^{\delta}(R) &= \CA\,\sum_{(ik) \in \Delta_{\delta}} \cC_{ik}\, \sum_{(\ell m) \in\Delta_\delta} \cC_{\ell m}\, \cG_{3,\mathrm{int}}^{(ik,\ell m)}(R), \\
\cG_{3,\mathrm{int}}^{(ik,\ell m)} &=\int_{1_\out} \int_{3_\inn}
\Bigg\{ \notag \\
&- \int_{2_\inn}  \left(\O_{12} \Ob_{13} \Ob_{23}\, \left[w_{ik}^1 \cA_{\ell m}^{23} + w_{ik}^2 \cA_{\ell m}^{13}\right]
+ \left(\O_{13} + \O_{12} \left(-1 + \Ob_{13} \Ob_{23}\right) \right) w_{ik}^3 \cA_{\ell m}^{12} \right) \notag \\
&+ \int_{2_\out} \left(\O_{13}\Ob_{23} w_{ik}^1 \cA_{\ell m}^{23} + \Ob_{13} \O_{23} \left[w_{ik}^2 \cA_{\ell m}^{13} + w_{ik}^3 \cA_{\ell m}^{12}\right]  \right) \Bigg\},
\end{align}
for the \textit{interference} part,
\begin{align}\label{eq:3loop:G3Dip}
\cG_{3,\mathrm{dip}}(R) &= \CAsq \sum_{(ik) \in \Delta_\delta} \cC_{ik}\,  \cG_{3,\mathrm{dip}}^{(ik)}(R), \\
\cG_{3,\mathrm{dip}}^{(ik)} &=\int_{1_\out} \int_{3_\inn} \Bigg\{ \int_{2_\inn} \left(\Ob_{13} \left(1 - \O_{12} \Ob_{23}\right) \cA_{ik}^{12} \cAb_{ik}^{13} -\O_{12} \Ob_{13} \Ob_{23} \cB_{ik}^{123} \right) \notag \\
&+ \int_{2_\out} \left(\Ob_{13} \O_{23}\, \cA_{ik}^{12} \cAb_{ik}^{13} - \Ob_{13} \Ob_{23}\, \cB_{ik}^{123} \right) \Bigg\},
\end{align}
for the \textit{dipole} part, and
\begin{align}\label{eq:3loop:G3Quad}
\cG_{3,\mathrm{quad}}^{\delta}(R) &= \sum_{(ik\ell) \in \pi_\delta} \cQ_\delta\,  \cG_{3,\mathrm{quad}}^{(ik\ell)}(R), \\
\cG_{3,\mathrm{quad}}^{(ik\ell)} &= \int_{1_\out} \int_{3_\inn} \Bigg\{ \int_{2_\inn} \Ob_{13} \left(1 - \O_{12} \Ob_{23}\right) \left[\cG_{ik}^{\ell 1}(2,3) + 2 \leftrightarrow 3 \right] \notag \\
&+ \int_{2_\out} \Ob_{13} \O_{23}\, \left[\cG_{ik}^{\ell 1}(2,3) + 2 \leftrightarrow 3 \right] \Bigg\},
\end{align}
\end{subequations}
for the \textit{quadrupole} part. Note there are 15 integrals to compute: $3^2$ for interference and 3 each for dipole and quadruple parts. The $a \leftrightarrow b$ symmetry reduces this to approximately 10 unique integrals.
Analytical evaluation was only feasible up to $\mathcal{O}(R^4)$ in the integrand's power series expansion (using the polar parametrisation \eqref{eq:Def:PolarParametrisation}). Beyond this order, analytical solutions proved intractable, necessitating numerical integration via the \texttt{Cuba} library. The resulting numerical output was fitted with a power series in $R$ up to $R^6$, with functional forms informed by anti-$k_t$ calculations from \cite{Khelifa-Kerfa:2024udm}.

\begin{figure}[t]
\centering
\includegraphics[scale=0.51]{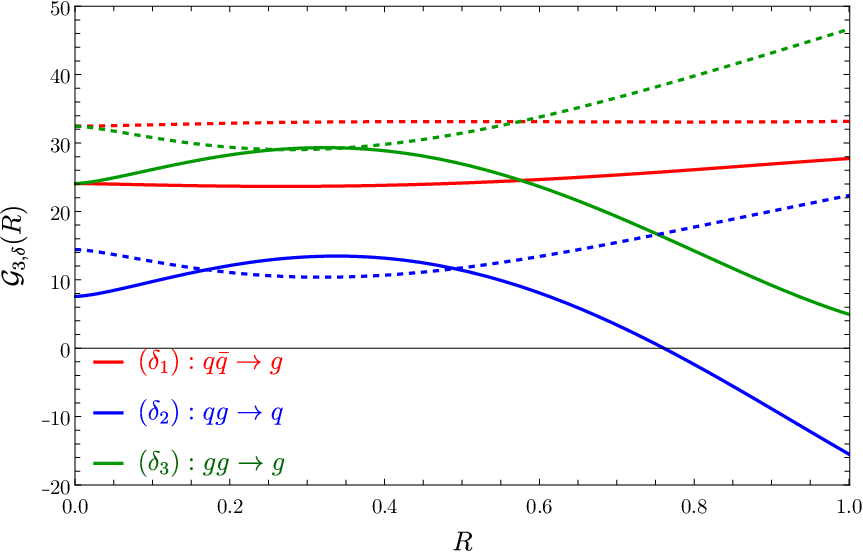}
\includegraphics[scale=0.51]{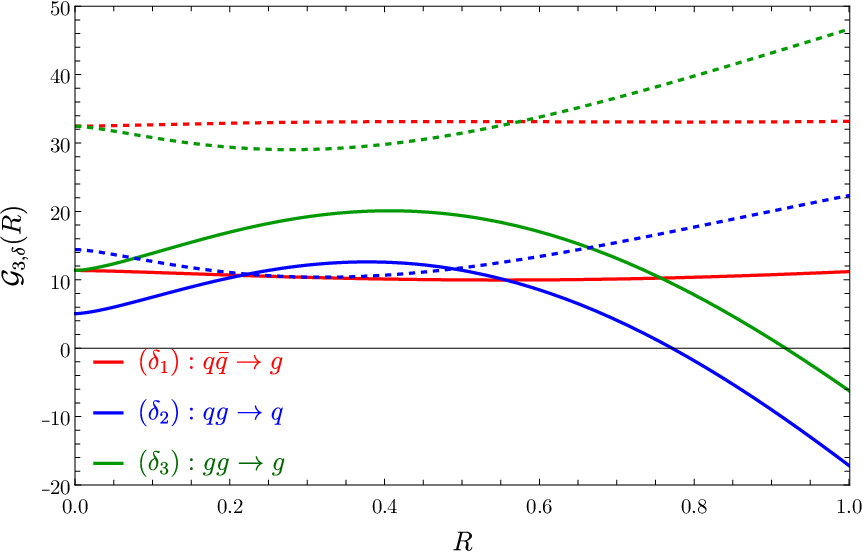}
\caption{Three-loop NGLs coefficients for $k_t$ (solid) and anti-$k_t$ (dashed) algorithms, showing full results (left) and interference-free contributions (right).}
\label{fig:3loop:G3}
\end{figure}

It is noteworthy that the integrand symmetries present in the anti-$k_t$ case for gluons 2 and 3 are absent for $k_t$ clustering. This stems from the latter algorithm's asymmetric treatment of gluons. For instance, in the quadrupole term \eqref{eq:3loop:G3Quad}, while the quadrupole antenna $\cG_{ik}^{\ell 1}(2,3)$ is symmetric under $2 \leftrightarrow 3$, the associated clustering factors lack this symmetry (e.g., $\Ob_{13} \O_{23} \neq \Ob_{12} \O_{23}$). The fitting expressions for the sum of all three parts in \eqref{eq:3loop:NGLs} for each channel are:
\begin{subequations}\label{eq:3loop:G3-channel}
\begin{align}\label{eq:3loop:G3-d1}
 \cG_{3,\delta_1}(R) &=
\CFsq \CA \left[0.002\,R^2 - \left(0.033 + 1.201 \ln R\right) R^4 + 0.286\,R^6  \right] + \notag\\
&+ \CF\CAsq \Big[-\left(1.559 + 1.259 \ln R - 2.423\ln^2 R\right) R^2 +\left(0.975 +6.090 \ln R\right) R^4 + \notag\\
&+ \left(-2.174 + 2.298\ln R\right) R^6 \Big]
+\CAcub \Big[0.891 + \left(0.738 +0.630 \ln R -1.121 \ln^2 R\right) R^2 - \notag\\
&-\left(0.484 + 2.745 \ln R \right) R^4 + \left(1.033 - 1.149 \ln R  \right) R^6  \Big] + \cO(R^8),
\end{align}
for channel $(\delta_1): q\bar{q} \to g$,
\begin{align}\label{eq:3loop:G3-d2}
\cG_{3,\delta_2}(R) &=
\CFsq \CA \left[0.471 +0.189\, R^2-0.013\, R^4 +0.0008\, R^6 \right] \notag\\
&+ \CF\CAsq \left[0.420 - \left(2.417 -0.287 \ln R \right) R^2 +\left(0.432 - 2.485 \ln R \right) R^4 + 0.824\, R^6  \right] \notag\\
&+\CAcub \Big[\left(1.411 -0.916 \ln R +1.211 \ln^2 R \right) R^2 + \left(0.039 +4.629 \ln R\right) R^4 \notag \\
& - \left(1.680 -1.149 \ln R\right) R^6 \Big] + \cO(R^8),
\end{align}
for channel $(\delta_2): qg \to q$, and
\begin{align}\label{eq:3loop:G3-d3}
\cG_{3,\delta_3}(R) &=
\CAcub \Big[0.891 + \left(0.059 - 0.630 \ln R + 1.211 \ln^2 R \right) R^2 + \left(0.351 + 3.387 \ln R \right) R^4 + \notag\\
&-\left(1.170 -1.149 \ln R \right) R^6 \Big]
+ \CA \Big[ 5.263\, R^2 - \left(0.641 - 7.454 \ln R \right) R^4 -1.887\, R^6 \Big] \notag\\ &+ \cO(R^8),
\end{align}
for channel $(\delta_3): gg \to g$.
\end{subequations}
Fig.~\ref{fig:3loop:G3} compares these expressions with their anti-$k_t$ counterparts from \cite{Khelifa-Kerfa:2024udm}\footnote{Note that \cite{Khelifa-Kerfa:2024udm} inadvertently swapped channels ($\delta_1$) and ($\delta_2$) due to misassignment of Born colour factors $\cC_{ik}$.}.
The following observations are to be noted:
\begin{enumerate}
\item The edge effect persists at three loops: NGLs coefficients approach non-zero constants as $R \to 0$ for all channels:
\begin{align}\label{eq:3loop:G3-R->0}
\lim_{R \to 0} \cG_{3,\delta_1} = \lim_{R \to 0} \cG_{3,\delta_3} &= 0.891\,\CAcub = \left(0.471 + 0.420\right) \CAcub, \notag \\
\lim_{R \to 0} \cG_{3,\delta_2} &= 0.471\,\CFsq \CA  + 0.420\, \CF \CAsq.
\end{align}
These match $e^+e^-$ results \cite{Khelifa-Kerfa:2024hwx} (eqs. (56) and (57) and figure 4B): $\CFsq \CA\, \lim_{R \to 0} \cG_{3,a}^{k_t} = 0.47\,\CFsq \CA$ and $\CF \CAsq \, \lim_{R \to 0} \cG_{3,b}^{k_t} = 0.42\,\CF\CAsq$ (identical to channel ($\delta_2$)). Channels ($\delta_1$) and ($\delta_3$) correspond to gluon-initiated jet analogues.

\item Unlike channel ($\delta_1$) where $k_t$ consistently reduces NGLs, channels ($\delta_2$) and ($\delta_3$) exhibit regions where $k_t$ coefficients match or exceed anti-$k_t$ values. For ($\delta_3$), this stems from the interference part \eqref{eq:3loop:G3Int} (absent in anti-$k_t$), evident in fig.~\ref{fig:3loop:G3}(right). For ($\delta_2$), the enhancement persists even without interference contributions.
This behaviour contrasts with established patterns at two loops and prior literature for $e^+e^-$ (up to four loops) and hadronic processes (up to two loops) \cite{Appleby:2002ke, Banfi:2005gj, Delenda:2006nf, Khelifa-Kerfa:2011quw, Kerfa:2012yae, Ziani:2021dxr, Khelifa-Kerfa:2024hwx}.

\item Anti-$k_t$ coefficients remain positive and $R$-monotonic, while $k_t$ coefficients exhibit channel-dependent behaviour:
\begin{itemize}
\item ($\delta_1$): Positive, slowly increasing with $R$ (similar to anti-$k_t$)
\item ($\delta_2$): Increases until $R \approx 0.4$, then decreases, turning negative for $R > 0.75$
\item ($\delta_3$): Similar shape to ($\delta_2$) but remains positive throughout
\end{itemize}
This structure arises because $k_t$ NGLs require the softest gluon (inside jet) to avoid being dragged out by harder gluons (outside jet). For larger jet radii $R$, the softest gluon is more likely to be geometrically closer to a harder gluon than to the jet-initiating parton. This proximity increases the probability of $k_t$ clustering between gluons rather than with the jet axis. Thus, configurations where the softest gluon avoids clustering despite this proximity become increasingly phase-space suppressed, leading to a reduction in the magnitude of the corresponding NGLs coefficient. For ($\delta_2$), the coefficient vanishes at $R \approx 0.77$.
\end{enumerate}

\begin{figure}[t]
\centering
\includegraphics[scale=0.7]{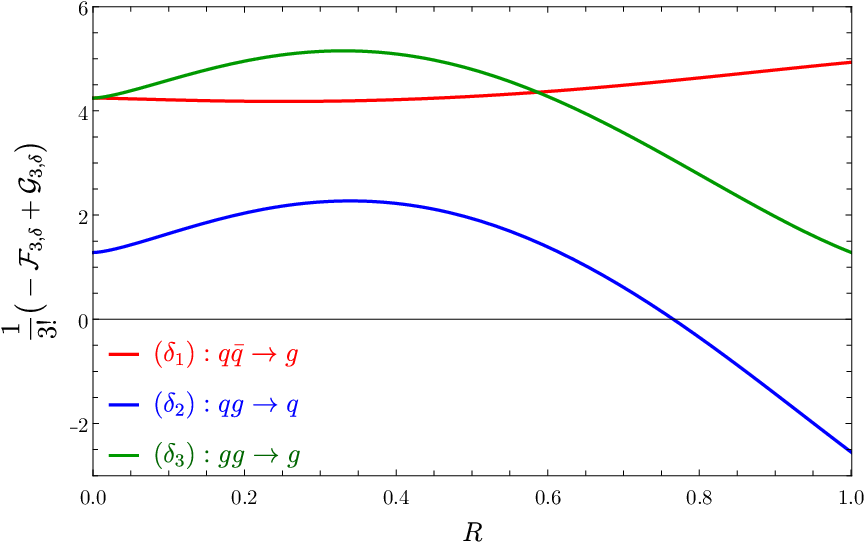}
\caption{Combined CLs and NGLs coefficients at three loops for $k_t$ clustering.}
\label{fig:3loop:F3-G3}
\end{figure}

The combined effect of CLs and NGLs at three loops is shown in fig.~\ref{fig:3loop:F3-G3} for all channels. CLs contributions are negligible compared to NGLs due to their smaller magnitude, leaving the $R$-dependence of NGLs coefficients largely unchanged. Moreover, since CLs coefficients are negative for all channels, the CLs distribution \eqref{eq:3loop:CLs} is positive and consequently amplifies rather than reduces the NGLs contribution. This compensation mechanism means the reduction in NGLs from $k_t$ clustering is partially offset by CLs enhancements. Thus, up to three loops, $k_t$ clustering cannot eliminate large logarithms originating from the non-global nature of the observable.

\subsection{Four-loops}
\label{sec:4loop}

For the emission of four soft, $k_t$-ordered gluons, the sum over all possible gluon configurations in the integrand of eq. \eqref{eq:FO:fB-mOrder} for a given partonic channel ($\delta$) is given by \cite{Khelifa-Kerfa:2024roc}:
\begin{align}\label{eq:4loop:uWX}
\sum_\X \Uh_4 \cW_{1234,\delta}^\X &= - \left(\prod_{i=1}^4 \Theta_i^\rho \right) \Theta_4^\inn
\Big[
                           \cW_{1234,\delta}^{\V\V\V\R}
 + \Theta_1^\out \Ob_{14}\,\cW_{1234,\delta}^{\R\V\V\R}
 + \Theta_2^\out \Ob_{24}\,\cW_{1234,\delta}^{\V\R\V\R} \notag\\
&+ \Theta_3^\out \Ob_{34}\,\cW_{1234,\delta}^{\V\V\R\R}
 + \Theta_1^\out \left(\Theta_2^\out + \Theta_2^\inn \O_{12} \right) \Ob_{14} \Ob_{24} \cW_{1234,\delta}^{\R\R\V\R} \notag\\
&+ \Theta_1^\out \left(\Theta_3^\out + \Theta_3^\inn \O_{13} \right) \Ob_{14} \Ob_{34} \cW_{1234,\delta}^{\R\V\R\R} \notag\\
&+ \Theta_2^\out \left(\Theta_3^\out + \Theta_3^\inn \O_{23} \right) \Ob_{24} \Ob_{34} \cW_{1234,\delta}^{\V\R\R\R} \notag\\
&+ \Theta_1^\out \left(\Theta_2^\out + \Theta_2^\inn \O_{12} \right) \left(\Theta_3^\out + \Theta_3^\inn \left[\O_{23} + \Ob_{23}\O_{13} \right] \right) \Ob_{14} \Ob_{24} \Ob_{34} \cW_{1234,\delta}^{\R\R\R\R}
\Big],
\end{align}
where the various components of the four-loop eikonal amplitude squared are defined in refs. \cite{Khelifa-Kerfa:2020nlc, Delenda:2015tbo}. Following an analogous procedure to the two- and three-loop cases, particularly through application of the complementarity relations $\Theta_i^\inn + \Theta_i^\out = 1$ and $\O_{ik} + \Ob_{ik} = 1$, we decompose this expression into contributions from three distinct phase-space regions.
These regions, excluding the softest gluon $k_4$ (always inside the jet, $\Theta_4^\inn$), comprise:
\begin{itemize}
\item Anti-$k_t$ region: $\Theta_1^\inn \Theta_2^\inn \Theta_3^\inn$ (no clustering contribution).
\item Interference regions: Three configurations yielding contributions from lower-order terms: $\Theta_1^\inn \Theta_2^\inn \Theta_3^\out$, $\Theta_1^\inn \Theta_2^\out \Theta_3^\inn$, and $\Theta_1^\inn \Theta_2^\out \Theta_3^\out$
\item New contribution regions: Four configurations generating new terms: $\Theta_1^\out \Theta_2^\out \Theta_3^\out$, $\Theta_1^\out \Theta_2^\out \Theta_3^\inn$, $\Theta_1^\out \Theta_2^\inn \Theta_3^\out$, and $\Theta_1^\out \Theta_2^\inn \Theta_3^\inn$
\end{itemize}
Upon integration, the jet mass cross-section adopts a form analogous to the two- \eqref{eq:2loop:fB} and three-loop \eqref{eq:3loop:fB} cases:
\begin{align}\label{eq:4loop:fB}
f_{\cB,\delta}^{(4)}(\rho) &= \frac{1}{4!} \left[ f_{\cB,\delta}^{(1)} \right]^4 + f_{\cB,\delta}^{(1)} \times \left[ \cC_{3,\delta} + \cS_{3,\delta} \right] + \frac{1}{2!} \left[ f_{\cB,\delta}^{(1)} \right]^2 \times \left[\cC_{2,\delta} + \cS_{2,\delta} \right] \notag\\
&+ \frac{1}{2!} \left[ \cC_{2,\delta}\right]^2 + \frac{1}{2} \left[ \cS_{2,\delta} \right]^2 + \cC_{4,\delta} + \cS_{4,\delta},
\end{align}
where $\cC_{4,\delta}$ and $\cS_{4,\delta}$ denote new four-loop CLs and NGLs contributions. These correspond to the following reduction of eq. \eqref{eq:4loop:uWX} (excluding the prefactor $-\prod_{i=1}^4 \Theta_i^\rho \Theta_4^\inn$):
\begin{align}\label{eq:4loop:fB-new}
&\Theta_1^\out \Theta_2^\out \Theta_3^\out \Big( \cW_{1234,\delta}^{\V\V\V\R} + \Ob_{14} \cW_{1234,\delta}^{\R\V\V\R} + \Ob_{24} \cW_{1234,\delta}^{\V\R\V\R} + \Ob_{34} \cW_{1234,\delta}^{\V\V\R\R} + \Ob_{14} \Ob_{24} \cW_{1234,\delta}^{\R\R\V\R} \notag\\
&\hspace{6em}+ \Ob_{14} \Ob_{34} \cW_{1234,\delta}^{\R\V\R\R} + \Ob_{24} \Ob_{34} \cW_{1234,\delta}^{\V\R\R\R} + \Ob_{14} \Ob_{24} \Ob_{34} \cW_{1234,\delta}^{\R\R\R\R} \Big)
\notag \\
&+\Theta_1^\out \Theta_2^\out \Theta_3^\inn \Big(\cW_{1234,\delta}^{\V\V\V\R} + \Ob_{14} \cW_{1234,\delta}^{\R\V\V\R} + \Ob_{24} \cW_{1234,\delta}^{\V\R\V\R} + \O_{13} \Ob_{14} \Ob_{34} \cW_{1234,\delta}^{\R\V\R\R} \notag\\
&\hspace{6em}+ \O_{23} \Ob_{24} \Ob_{34} \cW_{1234,\delta}^{\V\R\R\R} + \Ob_{14} \Ob_{24} \cW_{1234,\delta}^{\R\R\V\R}
+\left(\O_{23} + \O_{13} \Ob_{23} \right) \Ob_{14} \Ob_{24} \Ob_{34} \cW_{1234,\delta}^{\R\R\R\R} \Big)
\notag \\
&+\Theta_1^\out \Theta_2^\inn \Theta_3^\out \Big(\cW_{1234,\delta}^{\V\V\V\R} + \Ob_{14} \cW_{1234,\delta}^{\R\V\V\R} + \Ob_{34} \cW_{1234,\delta}^{\V\V\R\R} + \Ob_{14} \Ob_{34} \cW_{1234,\delta}^{\R\V\R\R} + \O_{12} \Ob_{14} \Ob_{24} \cW_{1234,\delta}^{\R\R\V\R} \notag\\
&\hspace{6em}+ \O_{12} \Ob_{14} \Ob_{24} \Ob_{34} \cW_{1234,\delta}^{\R\R\R\R} \Big)
\notag\\
&+ \Theta_1^\out \Theta_2^\inn \Theta_3^\inn \Big(\cW_{1234,\delta}^{\V\V\V\R} + \Ob_{14} \cW_{1234,\delta}^{\R\V\V\R}
+ \O_{13} \Ob_{14} \Ob_{34} \cW_{1234,\delta}^{\R\V\R\R}
+ \O_{12} \Ob_{14} \Ob_{24} \cW_{1234,\delta}^{\R\R\V\R}
\notag\\
&\hspace{6em}+ \O_{12} \left(\O_{23} + \O_{13} \Ob_{23}\right) \Ob_{14} \Ob_{24} \Ob_{34} \cW_{1234,\delta}^{\R\R\R\R} \Big).
\end{align}
Substituting the explicit expressions for the eikonal amplitudes squared allows computation of the four-loop CLs and NGLs coefficients as functions of the jet radius $R$. We now address each contribution separately.

\subsubsection{CLs}

\begin{figure}[t!]
\centering
\includegraphics[scale=0.7]{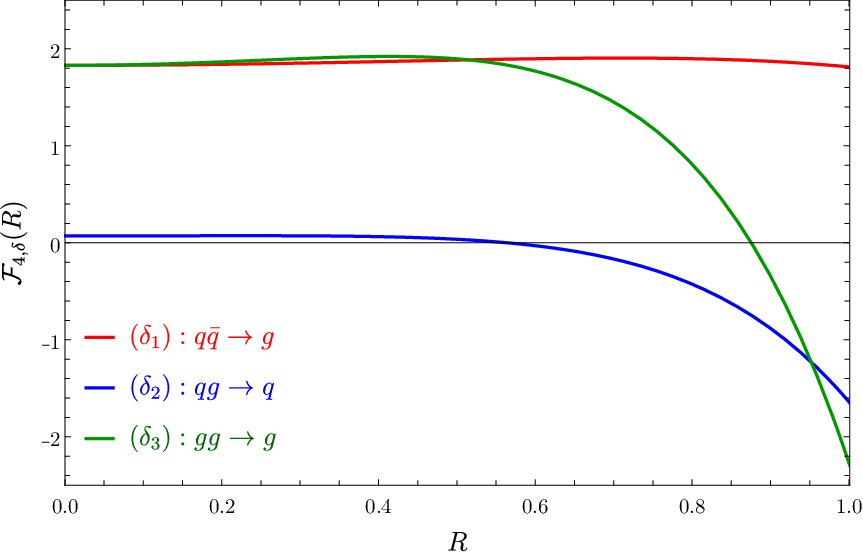}
\caption{The CLs coefficients at four-loops for each channel, for the $k_t$ jet algorithm.}
\label{fig:4loop:F4}
\end{figure}

Analogous to the two- and three-loop cases, the four-loop CLs contribution to the jet mass cross-section takes the form:
\begin{align}\label{eq:4loop:CLs}
 \cC_{4,\delta}(\rho) = + \frac{1}{4!}\,\asb^4\, L^4\, \cF_{4,\delta}(R),
\end{align}
where the four-loop CLs coefficient (see ref. \cite{Khelifa-Kerfa:2024hwx} for its $e^+e^-$ analogue) is:
\begin{align}\label{eq:4loop:F4-delta}
 \cF_{4,\delta}(R) &= \sum_{(ik) \in \Delta_\delta} \sum_{(\ell m) \in \Delta_\delta}  \sum_{(ns) \in \Delta_\delta}  \sum_{(qr) \in \Delta_\delta} \cC_{ik} \cC_{\ell m} \cC_{ns} \cC_{qr}\, R^8\, \cF_{4}^{(ik,\ell m,ns,qr)},
\end{align}
with
\begin{align}\label{eq:4loop:F4}
\cF_{4}^{(ik,\ell m,ns,qr)} &= \Bigg[
	\int_{1_\out} \int_{2_\out} \int_{3_\out} \int_{4_\inn} \O_{14} \O_{24} \O_{34}
	+\int_{1_\out} \int_{2_\out} \int_{3_\inn} \int_{4_\inn}
	\notag\\&\quad\times \Big[\O_{13} \O_{24} \left(-1+\Ob_{14} \Ob_{34} \right) + \O_{23} \left(-\O_{13} -\O_{14} +\Ob_{24} \Ob_{34} \left(1-\Ob_{13} \Ob_{14}\right) \right) \Big]
	\notag\\
	&+\int_{1_\out} \int_{2_\inn} \int_{3_\out} \int_{4_\inn} \O_{12} \O_{34} \left(-1+\Ob_{14} \Ob_{24}\right)
	\notag\\
	&+ \int_{1_\out} \int_{2_\inn} \int_{3_\inn} \int_{4_\inn} \O_{12} \left(-1+\Ob_{13} \Ob_{23}\right) \left(-1+\Ob_{14} \Ob_{24} \Ob_{34}\right) \Bigg] w_{ik}^1 w_{\ell m}^2 w_{ns}^3 w_{qr}^4.
\end{align}
The integrals were evaluated numerically, with results shown in fig.~\ref{fig:4loop:F4}.
The numerical results were fitted using a power series in $R$, yielding:
\begin{subequations}\label{eq:4loop:F4-FittingFuns}
\begin{align}
 \cF_{4,\delta_1}(R) &= 1.829 + 0.280\,R^2 - 0.221\,R^4 - 0.074\,R^6 + \cO(R^8), \\
 \cF_{4,\delta_2}(R) &= 0.071 + 0.063\,R^2 - 0.493 \,R^4 - 1.281\,R^6 + \cO(R^8), \\
 \cF_{4,\delta_3}(R) &= 1.829 + 0.997\,R^2 - 2.176\,R^4 - 2.920\,R^6 + \cO(R^8).
\end{align}
\end{subequations}
Notably, while the CLs coefficient for channel ($\delta_1$) remains positive throughout the $R$-range, those for channels ($\delta_2$) and ($\delta_3$) exhibit sign changes. Consequently, the CLs contribution to the jet mass in eq. \eqref{eq:4loop:CLs} is not uniformly positive. Specifically, fig.~\ref{fig:4loop:F4} shows that $\cF_{4,\delta_2}$ and $\cF_{4,\delta_3}$ vanish at $R = 0.56$ and $R = 0.87$ respectively. This sign-changing behaviour contrasts with the two- and three-loop results (Figs. \ref{fig:2loop:F2} and \ref{fig:3loop:F3}), where CLs coefficients maintain constant signs for $R \in [0,1]$.

Other characteristics observed at lower loops persist. In the vanishing jet-radius limit:
\begin{align}
\lim_{R \to 0} \cF_{4,\delta_1} = \lim_{R \to 0} \cF_{4,\delta_3} &= 1.83 = 0.0226\,\CAfour, \notag \\
 \lim_{R \to 0} \cF_{4,\delta_2} &= 0.071 = 0.0226\,\CFfour.
\end{align}
These results align with $e^+e^-$ findings \cite{Khelifa-Kerfa:2024hwx}: channel ($\delta_2$) matches identically, while channels ($\delta_1$) and ($\delta_3$) provide the gluon-jet analogues.

\subsubsection{NGLs}

\begin{figure}[t!]
\centering
\includegraphics[scale=0.7]{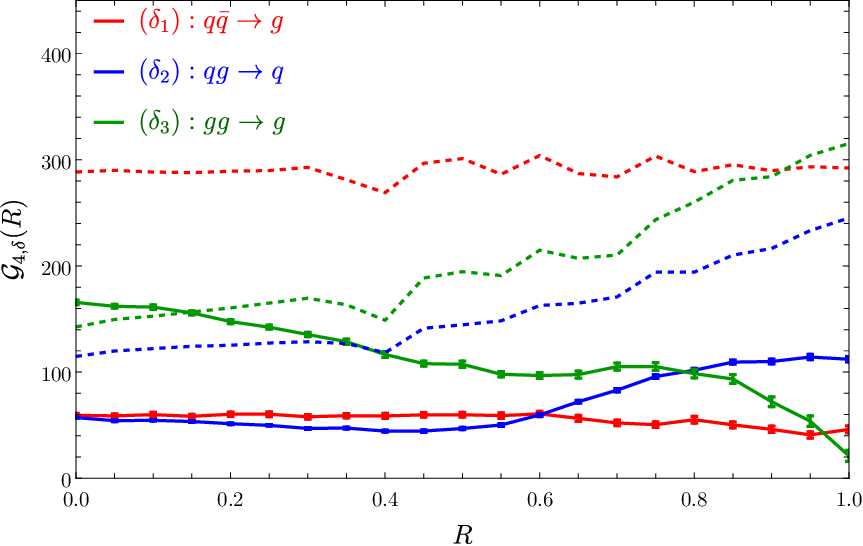}
\caption{The NGLs coefficients at four loops for each channel, for the $k_t$ (solid lines) and anti-$k_t$ (dashed lines) jet algorithms.}
\label{fig:4loop:G4}
\end{figure}

Substituting the explicit eikonal amplitudes squared from ref. \cite{Khelifa-Kerfa:2020nlc} into eq. \eqref{eq:4loop:fB-new}, we find that the NGLs contribution decomposes into three parts analogous to the three-loop case: interference, dipole, and quadrupole. The resulting expressions are cumbersome and thus omitted for brevity.
Note that deactivating clustering (setting all $\O_{ik} = 0$) reduces eq. \eqref{eq:4loop:fB-new} to the anti-$k_t$ form (eq. (3.26) in \cite{Khelifa-Kerfa:2024udm}). The four-loop NGLs contribution to the jet mass cross-section follows the same functional form as at two \eqref{eq:2loop:NGLs-B} and three loops \eqref{eq:3loop:NGLs}:
\begin{align}\label{eq:4loop:NGLs}
 \cS_{4,\delta}(\rho) = -\frac{1}{4!}\,\asb^4\,L^4\,\cG_{4,\delta}(R),
\end{align}
where $\cG_{4,\delta}$ denotes the four-loop NGLs coefficient for channel ($\delta$). Numerical integration results appear in fig.~\ref{fig:4loop:G4}, with each coefficient incorporating both colour and kinematic factors.
In the small-$R$ limit, the NGLs coefficients approach:
\begin{align}
\lim_{R \to 0} \cG_{4,\delta_1} = 59 \pm 1.71, \quad
\lim_{R \to 0} \cG_{4,\delta_2} = 57 \pm 0.53, \quad
\lim_{R \to 0} \cG_{4,\delta_3} = 165 \pm 2.10.
\end{align}
Comparison with $e^+e^-$ results \cite{Khelifa-Kerfa:2024hwx} ($\cG_4^{k_t} = -\CFcub \CA\, \cG_{4,a} - \CFsq \CAsq\,\cG_{4,b} +\CF \CAcub\,\cG_{4,c} + \CF \CAsq(\CA-2\CF)\, \cG_{4,d} = 50.79$) reveals discrepancies. Similar differences were noted for anti-$k_t$ in \cite{Khelifa-Kerfa:2024udm}. We attribute this to the four-loop quadrupole "ghost" term $\bar{\cN}_{1234}^{\R\R\V\R}$ within $\cW_{1234,\delta}^{\R\R\V\R}$ (see \cite{Khelifa-Kerfa:2020nlc} for details of its peculiar properties). Moreover, unlike the CLs case at two, three and four loops, channels ($\delta_1$) and ($\delta_3$) do not coincide at $R = 0$ for NGLs, likely due to ghost-term contributions from quadrupole colour factors.

Overall, the \(k_t\) algorithm yields smaller NGLs coefficients than anti-\(k_t\) across most values of \(R\). An exception arises in channel \((\delta_3)\), for which the \(k_t\) coefficient slightly exceeds its anti-\(k_t\) counterpart in the range \(0<R\lesssim0.15\). This behaviour mirrors the three-loop findings for channels \((\delta_2)\) and \((\delta_3)\), confirming that \(k_t\) clustering does not universally suppress NGLs at all radii. In the collinear limit \(R\to0\), \(k_t\) clustering reduces NGLs by approximately 70 \% in channel \((\delta_1)\) and by around 50 \% in channel \((\delta_2)\). At the phenomenologically relevant jet radius \(R=0.7\), the reductions become roughly 82 \% for \((\delta_1)\) and 50 \% for both \((\delta_2)\) and \((\delta_3)\). Consequently, for jet radii near 0.7, \(k_t\) clustering remains the preferable choice for minimising non-global logarithmic effects.

\begin{figure}[t!]
\centering
\includegraphics[scale=0.7]{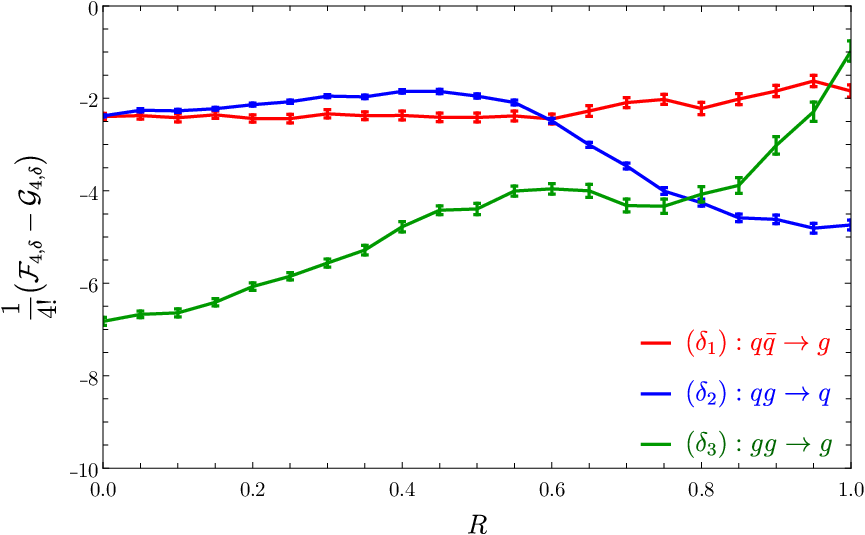}
\caption{The combined contribution of CLs and NGLs coefficients at four-loops for each channel, for the $k_t$ jet algorithms.}
\label{fig:4loop:G4-F4}
\end{figure}
Fig.~\ref{fig:4loop:G4-F4} shows the combined CLs and NGLs contributions at four loops as a function of jet radius $R$ for $k_t$ clustering. As observed at two and three loops, the partial cancellation between these logarithmic contributions does not eliminate their significance. Substantial net effects persist across all three channels for most $R$ values.

\section{Comparisons to all-orders results}
\label{sec:Comparisons}

The structure of the jet mass distribution at two, three and four loops (Eqs. \eqref{eq:2loop:fB}, \eqref{eq:3loop:fB} and \eqref{eq:4loop:fB}) suggests an exponential pattern. To assess the impact of higher-loop contributions relative to the standard two-loop result, we compare the exponentiated fixed-order results up to four loops with the output of the MC code from ref. \cite{Dasgupta:2001sh}. This code remains the only numerical implementation capable of resumming both NGLs and CLs for the $k_t$ algorithm, though limited to the large-$N_c$ approximation and single-logarithmic accuracy.
We parametrise its NGLs output as \cite{Dasgupta:2012hg,Ziani:2021dxr, Khelifa-Kerfa:2024udm}:
\begin{align}\label{eq:AllOrder:MCForm}
 \cS_\delta^{\MC}(t) = \exp\left[ -\CA \sum_{(ij) \in \Delta_\delta} \cC_{ij} \, \cG_{2}^{(ij)}\, f_{ij}(t) \right],
\end{align}
where $\Delta_{\delta}$ denotes the set of dipoles and $\cC_{ij}$ the dipole colour factors (previously defined in Sec. \ref{sec:FO}). The channel-specific two-loop NGLs coefficients $\cG_{2}^{(ij)}$ for dipoles $(ij)$ are given in eq. \eqref{eq:2loop:G2-B}.
The functional $f_{ij}$ and evolution variable $t$ are defined by:
\begin{align}\label{eq:AllOrder:fij-t}
 f_{ij}(t) = \frac{1 + \left(\lambda_{ij} t\right)^2}{1+ \left(\sigma_{ij} t\right)^{\gamma_{ij}}}\, t^2, \qquad
t = -\frac{1}{4 \pi \beta_0} \ln\left(1 - 2 \as(p_t) \beta_0 L \right),
\end{align}
where $L = \ln(R^2/\rho)$, $p_t$ is the measured jet's transverse momentum, and $\beta_0 = (11\CA - 2n_f)/(12\pi)$ is the leading QCD beta-function coefficient. The fitting parameters for $R = 0.7$ are:
\begin{align}\label{eq:AllOrder:FitParams}
\lambda_{aj} = \lambda_{bj} &= 2.29\CA, & \sigma_{aj} = \sigma_{bj} &= 10\CA, & \gamma_{aj} = \gamma_{bj} &= 0.88, \notag \\
\lambda_{ab} &= 1.12\CA, & \sigma_{ab} &= 4\CA, & \gamma_{ab} &= 0.68.
\end{align}
The analytical exponential for NGLs is:
\begin{align}\label{eq:AllOrder:St}
 \cS_\delta(t) = \exp\left[ -\sum_{n=2}^4 \frac{(-1)^n}{n!} \cG_{n,\delta}(R) (2 t)^n  \right],
\end{align}
noting that at fixed order $\asb L = 2 t$. The two-, three-, and four-loop NGLs coefficients appear in eqs. \eqref{eq:2loop:G2-channel-Final}, \eqref{eq:3loop:G3-channel}, and \eqref{eq:4loop:NGLs} (fig.~\ref{fig:4loop:G4}).
\begin{figure}[t!]
\centering
\includegraphics[scale=0.5]{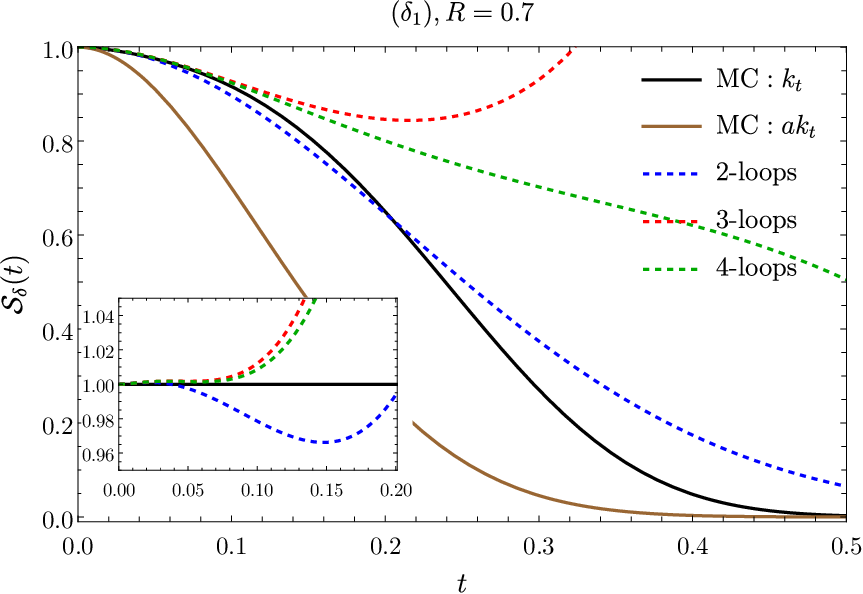}
\includegraphics[scale=0.5]{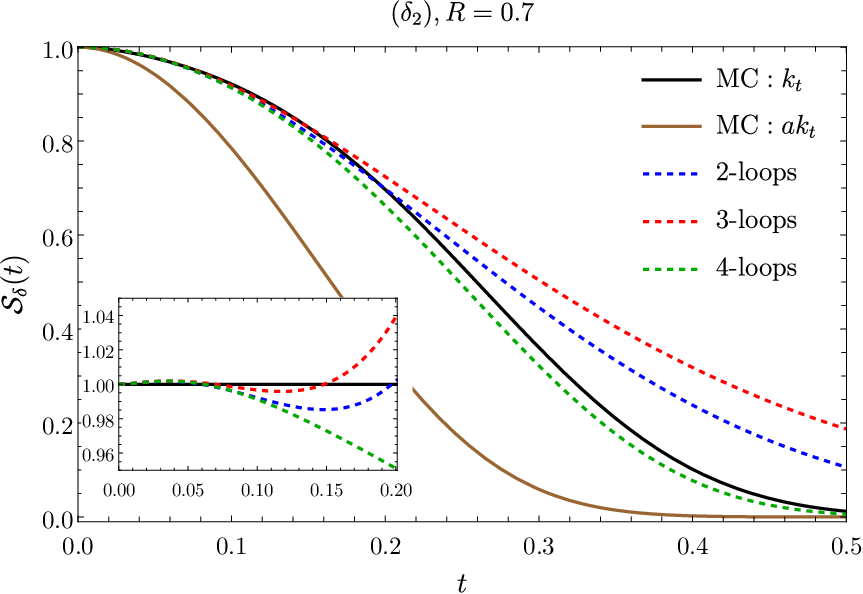}
\includegraphics[scale=0.5]{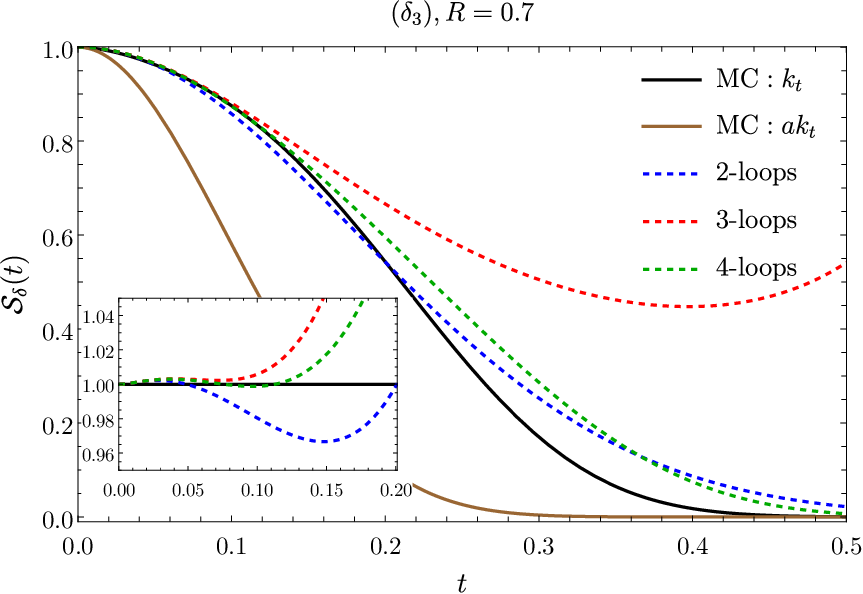}
\caption{Comparisons between the analytical exponentiation of the fixed-order NGLs coefficients and the output of the MC code of \cite{Dasgupta:2001sh} for $R = 0.7$, for all three channels. }
\label{fig:AllOrder:St}
\end{figure}

Figure \ref{fig:AllOrder:St} compares results for $R = 0.7$ across all partonic channels. Labels indicate truncation levels: "2-loops" (exponent up to $t^2$), "3-loops" ($t^3$), etc., with anti-$k_t$ MC results shown for reference. As established previously \cite{Ziani:2021dxr, Khelifa-Kerfa:2024udm}, the two-loop approximation reasonably describes the all-orders distribution for most $t$ values across all channels. Higher-loop effects become particularly evident at small $t$:
\begin{itemize}
\item Channel ($\delta_3$): Four-loop results match MC up to $t \sim 0.14$, performing comparably to two-loop for the remaining range of $t$.
\item Channel ($\delta_2$): Four-loop approximation outperforms lower orders across the full $t$-range.
\item Channel ($\delta_1$): Three- and four-loop results show best agreement at small $t$ ($t < 0.1$), while two-loop results show better agreement over the rest of the $t$-range.
\end{itemize}
Thus, clear improvements over two-loop accuracy appear: either at small $t$ (all channels) or across the full $t$-range ($\delta_2$). Note that $k_t$ clustering reduces NGLs impact relative to anti-$k_t$, as evidenced by the two MC curves.
\begin{figure}[t!]
\centering
\includegraphics[scale=0.5]{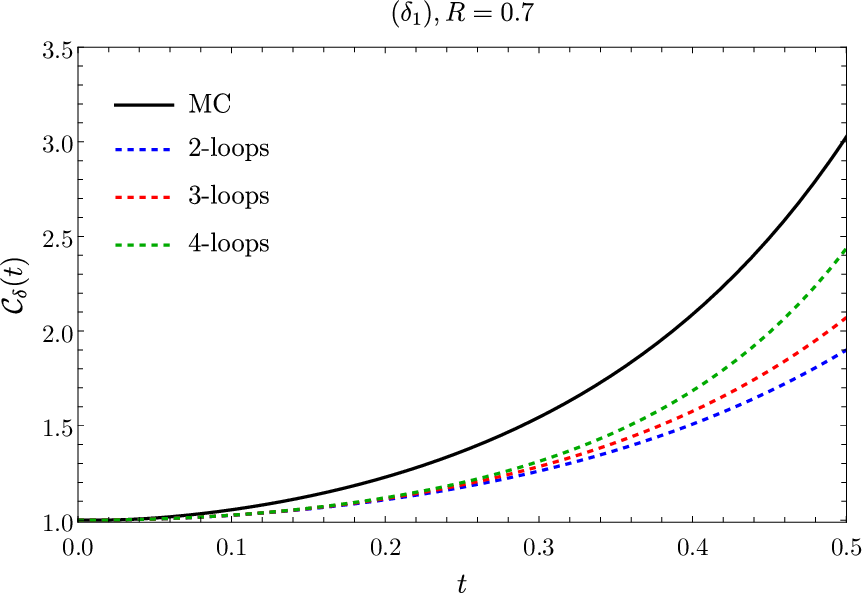}
\includegraphics[scale=0.5]{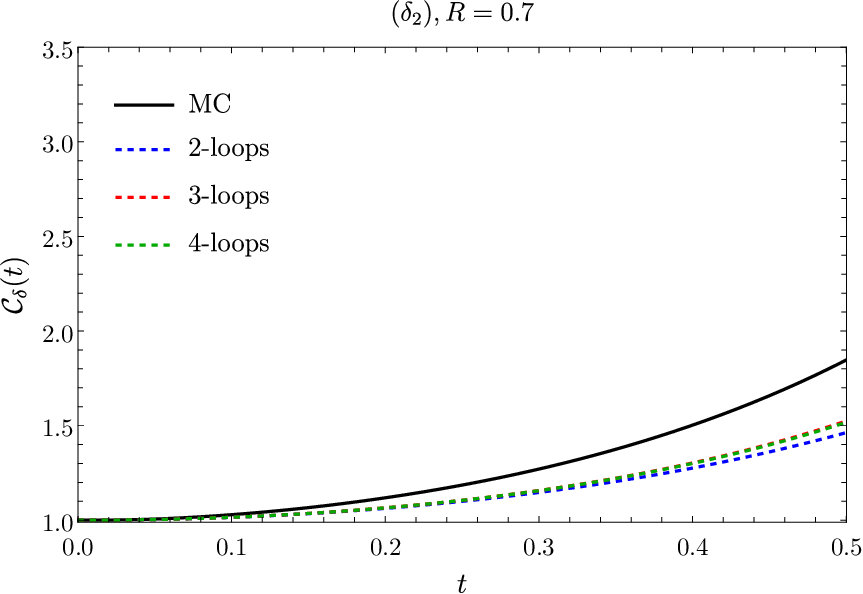}
\includegraphics[scale=0.5]{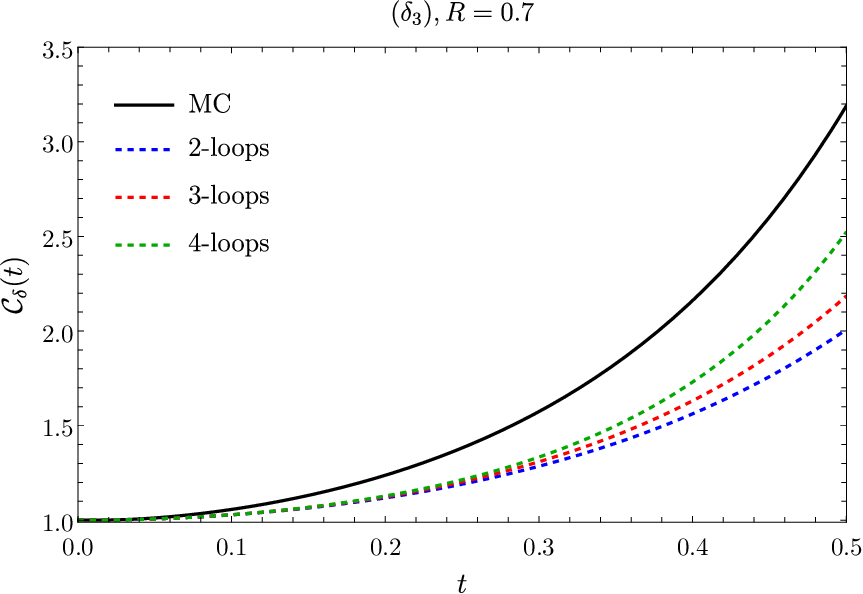}
\caption{Comparisons between the analytical exponentiation of the fixed-order CLs coefficients (dipole contribution only) and the output of the MC code of \cite{Dasgupta:2001sh} for $R = 0.7$, for all three channels.}
\label{fig:AllOrder:Ct}
\end{figure}

For CLs, the Monte Carlo code \cite{Dasgupta:2001sh} evolves only single dipoles, thus producing solely the \textit{dipole} contribution. Interference contributions require simultaneous evolution of multiple dipoles, which lies beyond the code's capabilities. Therefore, we compare only the dipole part of our analytical calculations to the MC output. We parametrise this output analogously to the NGLs case:
\begin{align}\label{eq:AllOrder:MCForm-CLs}
 \cC_\delta^{\MC}(t) = \exp\left[ \sum_{(ij) \in \Delta_\delta} \cC_{ij}^2 \, \cF_{2, \text{dip}}^{(ij)}\, f_{ij}(t) \right],
\end{align}
where $\cF_{2, \text{dip}}^{(ij)}$ are the two-loop dipole coefficients from eqs. \eqref{eq:2loop:F2-dipA} and \eqref{eq:2loop:F2-dipB}. The fitting parameters for $R = 0.7$ are:
\begin{align}\label{eq:AllOrder:FitParams-CLs}
\lambda_{aj} = \lambda_{bj} &= 19.71\,\CA, & \sigma_{aj} = \sigma_{bj} &= 8.43\,\CA, & \gamma_{aj} = \gamma_{bj} &= 2.18, \notag \\
\lambda_{ab} &= 3.34\,\CA, & \sigma_{ab} &= 2.14\,\CA, & \gamma_{ab} &= 2.18.
\end{align}
The analytical expression for comparison is:
\begin{align}\label{eq:AllOrder:Ct}
 \cC_\delta(t) = \exp\left[ \sum_{n=2}^4 \frac{(-1)^n}{n!} \cF_{n,\delta}(R) (2 t)^n  \right],
\end{align}
where the two-loop CLs coefficient $\cF_{2,\delta}(R)$ (from eq. \eqref{eq:2loop:F2-B}) includes only the dipole contribution. Fig.~\ref{fig:AllOrder:Ct} compares this analytical formula with the MC parametrisation \eqref{eq:AllOrder:MCForm-CLs} for $R=0.7$ across all channels. Truncation levels ("2-loops", "3-loops", etc.) match the NGLs convention in fig.~\ref{fig:AllOrder:St}.
Overall, higher-loop coefficients noticeably improve agreement, particularly for channels ($\delta_1$) and ($\delta_3$).

%
\begin{figure}[t!]
\centering
\includegraphics[scale=0.7]{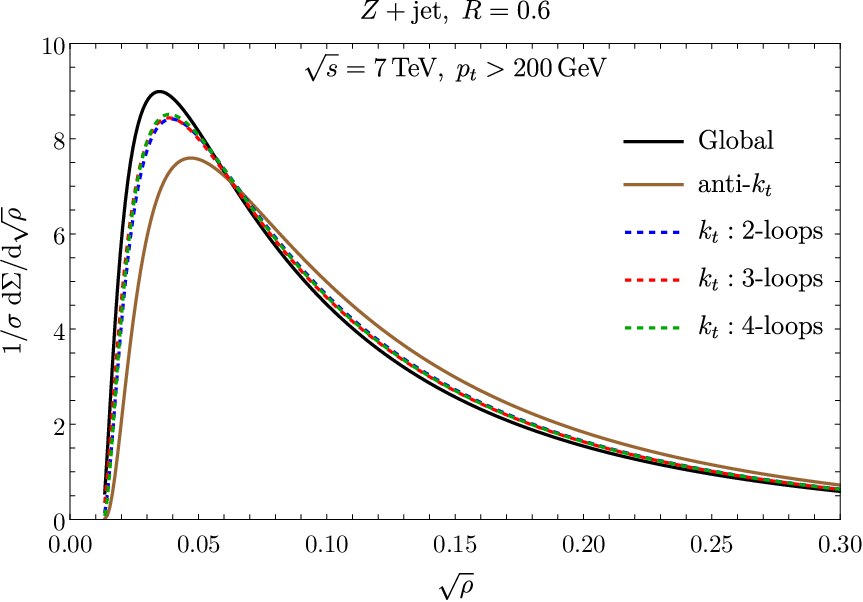}
\caption{Comparisons between the global distribution of the integrated jet mass cross section (summed up over all partonic channel) \eqref{eq:Def:IntegJetMassXsection} and the full resummed form factor that includes the effect of NGLs (at two-loops only) for the anti-$k_t$ and NGLs and CLs for the $k_t$ clustering at various loop orders. . }
\label{fig:AllOrder:Zj}
\end{figure}
To quantify the impact of higher-loop contributions on the total resummed form factor $\S(\rho)$ \eqref{eq:Def:IntegJetMassXsection}, fig.~\ref{fig:AllOrder:Zj} shows the derivative of $\S(\rho)$ with respect to $\sqrt{\rho}$ (square-root of invariant jet mass) for $Z$+jet production at $R = 0.6$ with fixed scales. The details of how such a plot may be produced, up to two-loops, are presented in section 6 of \cite{Ziani:2021dxr}.
The truncation scheme applies uniformly to both NGLs and CLs form factors (eqs. \eqref{eq:AllOrder:St} and \eqref{eq:AllOrder:Ct}): "2-loops" indicates truncation at second order, "3-loops" at third order, and "4-loops" at fourth order. Compared to the global-only distribution, anti-$k_t$ clustering reduces the peak region by approximately $15.5\%$ due to NGLs. This reduction diminishes progressively with $k_t$ clustering: $6.43\%$ at two loops, $6.15\%$ at three loops, and $5.35\%$ at four loops. Additionally, the $k_t$ distribution better approximates the tail region of the global distribution than anti-$k_t$.

The systematic reduction across loop orders suggests potential for further convergence toward the global distribution at higher loops. This convergence would require confirmation through higher-loop calculations, a task that will be addressed in the coming publications. This observation applies to the purely resummed distribution and may be modified by fixed-order matching and non-perturbative effects (see sections 6 and 7 of \cite{Ziani:2021dxr}).

\section{Conclusion}
\label{sec:Conclusion}

In this paper, we have extended our prior work on jet mass distributions in V/H+jet processes to the $k_t$ algorithm, providing the first comprehensive calculation of both clustering logarithms (CLs) and non-global logarithms (NGLs) up to four-loops in perturbative QCD. Building upon our two-loop results from a previous study~\cite{Ziani:2021dxr}, we derived semi-analytical expressions that maintain full dependence on colour factors and the jet radius $R$ for all relevant partonic channels. Our calculations were performed within the eikonal approximation, assuming strong energy ordering, which ensures single-logarithmic accuracy.

Our analysis revealed several key features and novel characteristics of $k_t$ clustering in the hadronic environment. We confirmed the ``edge effect'' where both CLs and NGLs coefficients remain non-zero as $R \to 0$, a phenomenon also observed in $e^+e^-$ processes. While the $k_t$ algorithm generally reduces NGLs compared to the anti-$k_t$ algorithm, consistent with lower-order studies~\cite{Khelifa-Kerfa:2024udm}, we identified new and unexpected behaviours at higher loop orders. Specifically, we found that for certain partonic channels, such as $gg \to gH$ with $0 < R \lesssim 0.15$, $k_t$ clustering can unexpectedly lead to larger NGLs than anti-$k_t$ clustering. This behaviour first emerges at three-loops and was therefore not present in our previous two-loop calculations. Furthermore, for the $qg \to qV/H$ channel, the three-loop NGLs coefficient becomes negative for $R > 0.75$, indicating a sign change in the logarithmic series. The CLs were found to partially compensate for the reduction in NGLs, particularly at larger values of $R$.

Our findings also underscored the dominance of gluon-initiated jets ($gg \to gH$ and $q\bar{q} \to gV/H$ channels) due to their larger colour factors. This dominance slightly increases with loop order, with the three- and four-loop NGLs coefficients for $gg \to gH$ being approximately three times larger than for quark-dominated channels at small $R$. Moreover, we observed that the $R$-dependence of the logarithmic coefficients is more pronounced for $k_t$ than for anti-$k_t$, with some coefficients changing sign across the $R$-range in specific channels. These particular features are absent in $e^+e^-$ annihilation processes and have not been previously reported.

The fixed-order perturbative series for both NGLs and CLs demonstrates reasonable convergence up to four-loops. The inclusion of these higher-loop terms improves agreement with all-orders resummation from numerical Monte Carlo methods, particularly at small values of the evolution variable $t$. We also assessed the impact of these higher-loop terms on the full resummed jet mass form factor in Z+jet production, finding that the reduction in the peak region of the distribution is less than 7\% at two-loops and becomes progressively smaller at three- and four-loop orders.

Several avenues for future research are apparent. These include extending the calculations to five loops to further text convergence and resummation frameworks, the inclusion of finite-$N_c$ corrections beyond the large-$N_c$ limit for Monte Carlo methods, the incorporation of subleading logarithms and recoil effects (which were neglected in this work), and the extension of these calculations to other jet algorithms such as Cambridge-Aachen and to more complex observables such as groomed jet mass.

\subsection*{Acknowledgment}

The authors extend their sincere gratitude to the Deanship of Scientific Research at the Islamic University of Madinah for its support of the Post‑Publishing Program.


\bibliographystyle{JHEP}
\bibliography{Refs}

\providecommand{\href}[2]{#2}\begingroup\raggedright\begin{thebibliography}{10}

\bibitem{Khelifa-Kerfa:2024udm}
K.~Khelifa-Kerfa, \emph{{Non-global logarithms up to four loops at
  finite-N$_{c}$ for V/H+jet processes at hadron colliders}},
  \href{https://doi.org/10.1007/JHEP10(2024)079}{\emph{JHEP} {\bfseries 10}
  (2024) 079} [\href{https://arxiv.org/abs/2406.13753}{{\ttfamily
  2406.13753}}].

\bibitem{Dasgupta:2001sh}
M.~Dasgupta and G.P.~Salam, \emph{{Resummation of nonglobal QCD observables}},
  \href{https://doi.org/10.1016/S0370-2693(01)00725-0}{\emph{Phys. Lett. B}
  {\bfseries 512} (2001) 323}
  [\href{https://arxiv.org/abs/hep-ph/0104277}{{\ttfamily hep-ph/0104277}}].

\bibitem{Banfi:2002hw}
A.~Banfi, G.~Marchesini and G.~Smye, \emph{{Away from jet energy flow}},
  \href{https://doi.org/10.1088/1126-6708/2002/08/006}{\emph{JHEP} {\bfseries
  08} (2002) 006} [\href{https://arxiv.org/abs/hep-ph/0206076}{{\ttfamily
  hep-ph/0206076}}].

\bibitem{Banfi:2010pa}
A.~Banfi, M.~Dasgupta, K.~Khelifa-Kerfa and S.~Marzani, \emph{{Non-global
  logarithms and jet algorithms in high-pT jet shapes}},
  \href{https://doi.org/10.1007/JHEP08(2010)064}{\emph{JHEP} {\bfseries 08}
  (2010) 064} [\href{https://arxiv.org/abs/1004.3483}{{\ttfamily 1004.3483}}].

\bibitem{Khelifa-Kerfa:2011quw}
K.~Khelifa-Kerfa, \emph{{Non-global logs and clustering impact on jet mass with
  a jet veto distribution}},
  \href{https://doi.org/10.1007/JHEP02(2012)072}{\emph{JHEP} {\bfseries 02}
  (2012) 072} [\href{https://arxiv.org/abs/1111.2016}{{\ttfamily 1111.2016}}].

\bibitem{Khelifa-Kerfa:2015mma}
K.~Khelifa-Kerfa and Y.~Delenda, \emph{{Non-global logarithms at finite N$_{c}$
  beyond leading order}},
  \href{https://doi.org/10.1007/JHEP03(2015)094}{\emph{JHEP} {\bfseries 03}
  (2015) 094} [\href{https://arxiv.org/abs/1501.00475}{{\ttfamily
  1501.00475}}].

\bibitem{Benslama:2020wib}
H.~Benslama, Y.~Delenda, K.~Khelifa-Kerfa and A.M.~Ibrahim, \emph{{Eikonal
  Amplitudes and Nonglobal Logarithms from the BMS Equation}},
  \href{https://doi.org/10.1134/S1547477121010039}{\emph{Phys. Part. Nucl.
  Lett.} {\bfseries 18} (2021) 5}
  [\href{https://arxiv.org/abs/2006.06738}{{\ttfamily 2006.06738}}].

\bibitem{Dasgupta:2012hg}
M.~Dasgupta, K.~Khelifa-Kerfa, S.~Marzani and M.~Spannowsky, \emph{{On jet mass
  distributions in Z+jet and dijet processes at the LHC}},
  \href{https://doi.org/10.1007/JHEP10(2012)126}{\emph{JHEP} {\bfseries 10}
  (2012) 126} [\href{https://arxiv.org/abs/1207.1640}{{\ttfamily 1207.1640}}].

\bibitem{Cacciari:2008gp}
M.~Cacciari, G.P.~Salam and G.~Soyez, \emph{{The anti-k$_t$ jet clustering
  algorithm}}, \href{https://doi.org/10.1088/1126-6708/2008/04/063}{\emph{JHEP}
  {\bfseries 04} (2008) 063} [\href{https://arxiv.org/abs/0802.1189}{{\ttfamily
  0802.1189}}].

\bibitem{Catani:1993hr}
S.~Catani, Y.L.~Dokshitzer, M.H.~Seymour and B.R.~Webber, \emph{{Longitudinally
  invariant $K_t$ clustering algorithms for hadron hadron collisions}},
  \href{https://doi.org/10.1016/0550-3213(93)90166-M}{\emph{Nucl. Phys. B}
  {\bfseries 406} (1993) 187}.

\bibitem{Ellis:1993tq}
S.D.~Ellis and D.E.~Soper, \emph{{Successive combination jet algorithm for
  hadron collisions}},
  \href{https://doi.org/10.1103/PhysRevD.48.3160}{\emph{Phys. Rev.} {\bfseries
  D48} (1993) 3160} [\href{https://arxiv.org/abs/hep-ph/9305266}{{\ttfamily
  hep-ph/9305266}}].

\bibitem{Ziani:2021dxr}
N.~Ziani, K.~Khelifa-Kerfa and Y.~Delenda, \emph{{Jet mass distribution in
  Higgs/vector boson + jet events at hadron colliders with $k_t$ clustering}},
  \href{https://doi.org/10.1140/epjc/s10052-021-09379-z}{\emph{Eur. Phys. J. C}
  {\bfseries 81} (2021) 570}
  [\href{https://arxiv.org/abs/2104.11060}{{\ttfamily 2104.11060}}].

\bibitem{Dokshitzer:1997in}
Y.L.~Dokshitzer, G.D.~Leder, S.~Moretti and B.R.~Webber, \emph{{Better jet
  clustering algorithms}},
  \href{https://doi.org/10.1088/1126-6708/1997/08/001}{\emph{JHEP} {\bfseries
  08} (1997) 001} [\href{https://arxiv.org/abs/hep-ph/9707323}{{\ttfamily
  hep-ph/9707323}}].

\bibitem{Wobisch:1998wt}
M.~Wobisch and T.~Wengler, \emph{{Hadronization corrections to jet
  cross-sections in deep inelastic scattering}},  in \emph{{Workshop on Monte
  Carlo Generators for HERA Physics (Plenary Starting Meeting)}}, pp.~270--279,
  4, 1998 [\href{https://arxiv.org/abs/hep-ph/9907280}{{\ttfamily
  hep-ph/9907280}}].

\bibitem{Khelifa-Kerfa:2024roc}
K.~Khelifa-Kerfa, \emph{{Analytical structure of kt clustering to any order}},
  \href{https://doi.org/10.1103/PhysRevD.111.014027}{\emph{Phys. Rev. D}
  {\bfseries 111} (2025) 014027}
  [\href{https://arxiv.org/abs/2409.14029}{{\ttfamily 2409.14029}}].

\bibitem{Appleby:2002ke}
R.B.~Appleby and M.H.~Seymour, \emph{{Nonglobal logarithms in interjet energy
  flow with k$_t$ clustering requirement}},
  \href{https://doi.org/10.1088/1126-6708/2002/12/063}{\emph{JHEP} {\bfseries
  12} (2002) 063} [\href{https://arxiv.org/abs/hep-ph/0211426}{{\ttfamily
  hep-ph/0211426}}].

\bibitem{Banfi:2005gj}
A.~Banfi and M.~Dasgupta, \emph{{Problems in resumming interjet energy flows
  with k(t) clustering}},
  \href{https://doi.org/10.1016/j.physletb.2005.08.125}{\emph{Phys. Lett.}
  {\bfseries B628} (2005) 49}
  [\href{https://arxiv.org/abs/hep-ph/0508159}{{\ttfamily hep-ph/0508159}}].

\bibitem{Delenda:2006nf}
Y.~Delenda, R.~Appleby, M.~Dasgupta and A.~Banfi, \emph{{On QCD resummation
  with k(t) clustering}},
  \href{https://doi.org/10.1088/1126-6708/2006/12/044}{\emph{JHEP} {\bfseries
  0612} (2006) 044} [\href{https://arxiv.org/abs/hep-ph/0610242}{{\ttfamily
  hep-ph/0610242}}].

\bibitem{Delenda:2012mm}
Y.~Delenda and K.~Khelifa-Kerfa, \emph{{On the resummation of clustering
  logarithms for non-global observables}},
  \href{https://doi.org/10.1007/JHEP09(2012)109}{\emph{JHEP} {\bfseries 09}
  (2012) 109} [\href{https://arxiv.org/abs/1207.4528}{{\ttfamily 1207.4528}}].

\bibitem{Bouaziz:2022tik}
H.~Bouaziz, Y.~Delenda and K.~Khelifa-Kerfa, \emph{{Azimuthal decorrelation
  between a jet and a Z boson at hadron colliders}},
  \href{https://doi.org/10.1007/JHEP10(2022)006}{\emph{JHEP} {\bfseries 10}
  (2022) 006} [\href{https://arxiv.org/abs/2207.10147}{{\ttfamily
  2207.10147}}].

\bibitem{Benslama:2023gys}
H.~Benslama, Y.~Delenda and K.~Khelifa-Kerfa, \emph{{Dijet azimuthal
  decorrelation in e+e\ensuremath{-} annihilation}},
  \href{https://doi.org/10.1016/j.physletb.2023.137903}{\emph{Phys. Lett. B}
  {\bfseries 840} (2023) 137903}
  [\href{https://arxiv.org/abs/2301.00860}{{\ttfamily 2301.00860}}].

\bibitem{Becher:2023znt}
T.~Becher and J.~Haag, \emph{{Factorization and resummation for sequential
  recombination jet cross sections}},
  \href{https://doi.org/10.1007/JHEP01(2024)155}{\emph{JHEP} {\bfseries 01}
  (2024) 155} [\href{https://arxiv.org/abs/2309.17355}{{\ttfamily
  2309.17355}}].

\bibitem{Khelifa-Kerfa:2025cdn}
K.~Khelifa-Kerfa and M.~Benghanem, \emph{{Hemisphere mass up to four-loops with
  generalised $k_t$ algorithms}},
  \href{https://arxiv.org/abs/2506.14415}{{\ttfamily 2506.14415}}.

\bibitem{Khelifa-Kerfa:2024hwx}
K.~Khelifa-Kerfa, \emph{{Dijet mass up to four loops with(out) $k_t$
  clustering}},
  \href{https://doi.org/10.1140/epjc/s10052-025-13837-3}{\emph{Eur. Phys. J. C}
  {\bfseries 85} (2025) 133}
  [\href{https://arxiv.org/abs/2411.03956}{{\ttfamily 2411.03956}}].

\bibitem{Khelifa-Kerfa:2024gyv}
K.~Khelifa-Kerfa, \emph{{Clustering logarithms up to six loops}},
  \href{https://arxiv.org/abs/2412.03244}{{\ttfamily 2412.03244}}.

\bibitem{Anastasiou:2016cez}
C.~Anastasiou, C.~Duhr, F.~Dulat, E.~Furlan, T.~Gehrmann, F.~Herzog et~al.,
  \emph{{High precision determination of the gluon fusion Higgs boson
  cross-section at the LHC}},
  \href{https://doi.org/10.1007/JHEP05(2016)058}{\emph{JHEP} {\bfseries 05}
  (2016) 058} [\href{https://arxiv.org/abs/1602.00695}{{\ttfamily
  1602.00695}}].

\bibitem{Mistlberger:2018etf}
B.~Mistlberger, \emph{{Higgs boson production at hadron colliders at N$^{3}$LO
  in QCD}}, \href{https://doi.org/10.1007/JHEP05(2018)028}{\emph{JHEP}
  {\bfseries 05} (2018) 028}
  [\href{https://arxiv.org/abs/1802.00833}{{\ttfamily 1802.00833}}].

\bibitem{Chen:2019fhs}
L.-B.~Chen, H.T.~Li, H.-S.~Shao and J.~Wang, \emph{{The gluon-fusion production
  of Higgs boson pair: N$^3$LO QCD corrections and top-quark mass effects}},
  \href{https://doi.org/10.1007/JHEP03(2020)072}{\emph{JHEP} {\bfseries 03}
  (2020) 072} [\href{https://arxiv.org/abs/1912.13001}{{\ttfamily
  1912.13001}}].

\bibitem{Chen:2025utl}
X.~Chen, X.~Guan and B.~Mistlberger, \emph{{Three-Loop QCD corrections to the
  production of a Higgs boson and a Jet}},
  \href{https://arxiv.org/abs/2504.06490}{{\ttfamily 2504.06490}}.

\bibitem{Gauld:2021ule}
R.~Gauld, A.~Gehrmann-De~Ridder, E.W.N.~Glover, A.~Huss and I.~Majer, \emph{{VH
  + jet production in hadron-hadron collisions up to order $
  {\alpha}_{\mathrm{s}}^3 $ in perturbative QCD}},
  \href{https://doi.org/10.1007/JHEP03(2022)008}{\emph{JHEP} {\bfseries 03}
  (2022) 008} [\href{https://arxiv.org/abs/2110.12992}{{\ttfamily
  2110.12992}}].

\bibitem{Boughezal:2015aha}
R.~Boughezal, C.~Focke, W.~Giele, X.~Liu and F.~Petriello, \emph{{Higgs boson
  production in association with a jet at NNLO using jettiness subtraction}},
  \href{https://doi.org/10.1016/j.physletb.2015.06.055}{\emph{Phys. Lett. B}
  {\bfseries 748} (2015) 5} [\href{https://arxiv.org/abs/1505.03893}{{\ttfamily
  1505.03893}}].

\bibitem{Boughezal:2015ded}
R.~Boughezal, J.M.~Campbell, R.K.~Ellis, C.~Focke, W.T.~Giele, X.~Liu et~al.,
  \emph{{Z-boson production in association with a jet at
  next-to-next-to-leading order in perturbative QCD}},
  \href{https://doi.org/10.1103/PhysRevLett.116.152001}{\emph{Phys. Rev. Lett.}
  {\bfseries 116} (2016) 152001}
  [\href{https://arxiv.org/abs/1512.01291}{{\ttfamily 1512.01291}}].

\bibitem{Caola:2015wna}
F.~Caola, K.~Melnikov and M.~Schulze, \emph{{Fiducial cross sections for Higgs
  boson production in association with a jet at next-to-next-to-leading order
  in QCD}}, \href{https://doi.org/10.1103/PhysRevD.92.074032}{\emph{Phys. Rev.
  D} {\bfseries 92} (2015) 074032}
  [\href{https://arxiv.org/abs/1508.02684}{{\ttfamily 1508.02684}}].

\bibitem{Gehrmann-DeRidder:2015wbt}
A.~Gehrmann-De~Ridder, T.~Gehrmann, E.W.N.~Glover, A.~Huss and T.A.~Morgan,
  \emph{{Precise QCD predictions for the production of a Z boson in association
  with a hadronic jet}},
  \href{https://doi.org/10.1103/PhysRevLett.117.022001}{\emph{Phys. Rev. Lett.}
  {\bfseries 117} (2016) 022001}
  [\href{https://arxiv.org/abs/1507.02850}{{\ttfamily 1507.02850}}].

\bibitem{Boughezal:2016dtm}
R.~Boughezal, X.~Liu and F.~Petriello, \emph{{W-boson plus jet differential
  distributions at NNLO in QCD}},
  \href{https://doi.org/10.1103/PhysRevD.94.113009}{\emph{Phys. Rev. D}
  {\bfseries 94} (2016) 113009}
  [\href{https://arxiv.org/abs/1602.06965}{{\ttfamily 1602.06965}}].

\bibitem{Gehrmann-DeRidder:2016cdi}
A.~Gehrmann-De~Ridder, T.~Gehrmann, E.W.N.~Glover, A.~Huss and T.A.~Morgan,
  \emph{{The NNLO QCD corrections to Z boson production at large transverse
  momentum}}, \href{https://doi.org/10.1007/JHEP07(2016)133}{\emph{JHEP}
  {\bfseries 07} (2016) 133}
  [\href{https://arxiv.org/abs/1605.04295}{{\ttfamily 1605.04295}}].

\bibitem{Campbell:2017dqk}
J.M.~Campbell, R.K.~Ellis and C.~Williams, \emph{{Driving missing data at the
  LHC: NNLO predictions for the ratio of $\gamma+j$ and $Z+j$}},
  \href{https://doi.org/10.1103/PhysRevD.96.014037}{\emph{Phys. Rev. D}
  {\bfseries 96} (2017) 014037}
  [\href{https://arxiv.org/abs/1703.10109}{{\ttfamily 1703.10109}}].

\bibitem{NNLOJET:2025rno}
{\scshape NNLOJET} collaboration, \emph{{NNLOJET: a parton-level event
  generator for jet cross sections at NNLO QCD accuracy}},
  \href{https://arxiv.org/abs/2503.22804}{{\ttfamily 2503.22804}}.

\bibitem{Barontini:2024eii}
A.~Barontini, N.~Laurenti and J.~Rojo, \emph{{NNPDF progress and the path to
  proton structure at N$^3$LO accuracy}},
  \href{https://doi.org/10.22323/1.469.0039}{\emph{PoS} {\bfseries DIS2024}
  (2025) 039}.

\bibitem{Banfi:2004yd}
A.~Banfi, G.P.~Salam and G.~Zanderighi, \emph{{Principles of general
  final-state resummation and automated implementation}},
  \href{https://doi.org/10.1088/1126-6708/2005/03/073}{\emph{JHEP} {\bfseries
  03} (2005) 073} [\href{https://arxiv.org/abs/hep-ph/0407286}{{\ttfamily
  hep-ph/0407286}}].

\bibitem{Cacciari:2011ma}
M.~Cacciari, G.P.~Salam and G.~Soyez, \emph{{FastJet User Manual}},
  \href{https://doi.org/10.1140/epjc/s10052-012-1896-2}{\emph{Eur. Phys. J. C}
  {\bfseries 72} (2012) 1896}
  [\href{https://arxiv.org/abs/1111.6097}{{\ttfamily 1111.6097}}].

\bibitem{Schwartz:2014wha}
M.D.~Schwartz and H.X.~Zhu, \emph{{Nonglobal logarithms at three loops, four
  loops, five loops, and beyond}},
  \href{https://doi.org/10.1103/PhysRevD.90.065004}{\emph{Phys. Rev. D}
  {\bfseries 90} (2014) 065004}
  [\href{https://arxiv.org/abs/1403.4949}{{\ttfamily 1403.4949}}].

\bibitem{Khelifa-Kerfa:2020nlc}
K.~Khelifa-Kerfa and Y.~Delenda, \emph{{Eikonal amplitudes for three-hard legs
  processes at finite-N$_c$}},
  \href{https://doi.org/10.1016/j.physletb.2020.135768}{\emph{Phys. Lett. B}
  {\bfseries 809} (2020) 135768}
  [\href{https://arxiv.org/abs/2006.08758}{{\ttfamily 2006.08758}}].

\bibitem{Delenda:2015tbo}
Y.~Delenda and K.~Khelifa-Kerfa, \emph{{Eikonal gluon bremsstrahlung at finite
  $N_c$ beyond two loops}},
  \href{https://doi.org/10.1103/PhysRevD.93.054027}{\emph{Phys. Rev. D}
  {\bfseries 93} (2016) 054027}
  [\href{https://arxiv.org/abs/1512.05401}{{\ttfamily 1512.05401}}].

\bibitem{Hahn:2004fe}
T.~Hahn, \emph{{CUBA: A Library for multidimensional numerical integration}},
  \href{https://doi.org/10.1016/j.cpc.2005.01.010}{\emph{Comput. Phys. Commun.}
  {\bfseries 168} (2005) 78}
  [\href{https://arxiv.org/abs/hep-ph/0404043}{{\ttfamily hep-ph/0404043}}].

\bibitem{Kerfa:2012yae}
K.~Khelifa-Kerfa, \emph{{QCD resummation for high-$p_T$ jet shapes at hadron
  colliders}}, Ph.D. thesis, Manchester U., 2012.
\newblock \href{https://arxiv.org/abs/2111.10671}{{\ttfamily 2111.10671}}.

\end{thebibliography}\endgroup
\end{document}